\documentclass[12pt,hidelinks]{article}
\usepackage{bm}
\usepackage{bbm}
\usepackage[semicolon]{natbib}
\usepackage[pdftex]{graphicx}
\DeclareGraphicsExtensions{.png,.pdf,.jpg}
\usepackage[hmargin=1in,vmargin=1in]{geometry}
\usepackage{setspace}
\usepackage{verbatim}
\usepackage{amssymb, amsthm}
\usepackage{amsmath}
\usepackage{booktabs}
\usepackage{hyperref}
\usepackage{appendix}
\usepackage{tcolorbox}
\usepackage{xcolor}
\usepackage{longtable}
\usepackage{afterpage}
\usepackage{pdflscape}
\usepackage{multirow}
\usepackage{capt-of}
\usepackage{lscape}
\usepackage{subcaption}
\usepackage{mathtools}
\usepackage[maxfloats=25]{morefloats}
\usepackage[export]{adjustbox}

\DeclareRobustCommand\sampleline[1]{%
  \tikz\draw[#1] (0,0) (0,\the\dimexpr\fontdimen22\textfont2\relax)
  -- (1em,\the\dimexpr\fontdimen22\textfont2\relax);%
}

\usepackage{float}

\theoremstyle{remark} 
\usepackage{tikz}
\usetikzlibrary{fit,positioning}

\theoremstyle{plain}

\newtheorem{proposition}{Proposition}
\theoremstyle{definition}

\theoremstyle{remark}

\title{A Bayesian Approach for Inference on Mixed Graphical Models}

\author{Mauro Florez\thanks{Department of Statistics, Rice University, USA}$~^\dag$, Anna Gottard\thanks{Department of Statistics, Computer Science, Application "G.Parenti", University of Florence, Italy}, Carrie McAdams\thanks{Department of Psychiatry, University of Texas at Southwestern Medical School, Dallas, TX, USA}, Michele Guindani\thanks{Department of Biostatistics,  University of California at Los Angeles, USA}\\ 
and\\
 Marina Vannucci$^*$}

\begin{document}
\maketitle

\begin{abstract}
Mixed data refers to a type of data in which variables can be of multiple types, such as continuous, discrete, or categorical. This data is routinely collected in various fields, including healthcare and social sciences. A common goal in the analysis of such data is to identify dependence relationships between variables, for an understanding of their associations. In this paper, we propose a Bayesian pairwise graphical model that estimates conditional independencies between any type of data. We implement a flexible modeling construction, that includes zero-inflated count data and can also handle missing data. We show that the model maintains both global and local Markov properties. We employ a spike-and-slab prior for the estimation of the graph and implement an MCMC algorithm for posterior inference based on conditional likelihoods.  We assess performances on four simulation scenarios with distinct dependence structures, that also include cases with data missing at random, and compare results with existing methods. Finally, we present an analysis of real data from adolescents diagnosed with an eating disorder. Estimated graphs show differences in the associations estimated at intake and discharge, suggesting possible effects of the treatment on cognitive and behavioral measures in the adolescents.
\end{abstract}

\noindent {\bf Keywords:} Eating disorders, Markov chain Monte Carlo, Missing data, Mixed data, Pairwise graphical model, Spike-and-slab prior

\section{Introduction}
Graphical models are a class of multivariate distributions whose conditional independence structure is encoded by a graph. Each node in the graph corresponds to a random variable, and missing edges indicate some conditional independence between the corresponding variables.
In particular, undirected graphs are characterized by undirected edges, 
ensuring a symmetric structure among variables. For continuous random variables that adhere to a multivariate Gaussian distribution, these models are often referred to as concentration graph models. They have been extensively studied, particularly in a high-dimensional setting
using estimators such as the graphical lasso (GLASSO)  \citep{friedman2008sparse, yuan2007model}, which combines parameter estimation and graph structural learning in a single step. Other approaches to Gaussian data include the neighborhood selection \citep{Meinshausen_Buhlmann_2006} and penalized likelihood methods \citep{Fan_Feng_Wu_2009}. In the Bayesian framework, models have been proposed based on the G-Wishart prior for the precision matrix \citep{dobra2011bayesian, lenkoski2011computational} or based on shrinkage priors, such as the Bayesian lasso \citep{Peterson_Vannucci_Karakas_Choi_Ma_Maletic-Savatic_2013}, the graphical horseshoe prior introduced by \citet{horseshoe_li}, and the continuous spike-and-slab prior proposed by \citet{wang2015scaling}. See \citet{Ni_Baladandayuthapani_Vannucci_Stingo_2022} for a recent review paper.

In the domain of categorical variables, common approaches to graph estimation have traditionally relied on classical log-linear models \citep{lauritzen1996graphical}. More recently, there has been a growing interest in the Ising model for multiple Bernoulli variables \citep{ravikumar2010high, lee2006efficient}, as well as in models where all node-conditional distributions belong to a single type, such as multinomial \citep{jalali2011learning}, Poisson \citep{yang2013poisson, allen2013local2} and other discrete data types \citep{Hue_Chiogna_2021}, or any univariate distribution in the exponential family \citep{yang2015graphical}. For count data, these approaches impose some restrictions to the parameter space, allowing
to capture only negative relationships between variables, with the exception of \citet{allen2013local2} and \citet{Hue_Chiogna_2021} who define node-conditional Poisson distributions without requiring a joint specification, enabling estimation of more flexible dependence structures. Bayesian approaches have also been proposed. For Ising models, \citet{park2022} developed an approach based on a spike-and-slab prior and applied it to analyze educational data. For count data, \citet{roy2020nonparametric} proposed a new class of pairwise graphical models, able to estimate both positive and negative interactions.

In this paper, we consider the framework of mixed data, which refers to a type of data in which the observed variables can be of multiple types, such as continuous, discrete, or categorical. This data is routinely collected in various fields, including healthcare and social sciences. For example, in the application of this paper we consider data on eating disorder symptoms, collected from adolescents admitted for treatment at a pediatric eating disorder center, which includes continuous, categorical and zero-inflated count variables.
%In many real data applications, datasets may contain mixed types of data, including continuous, discrete, categorical, or zero-inflated data. 
For mixed data, \citet{yang2015mixed} proposed a frequentist mixed graphical model where the node-conditional distributions belong to the exponential family. However, they provide conditions for normalizability only in the case of graphical models with two types of nodes, such as Gaussian-Poisson, Poisson-Ising, or Gaussian-Ising. Later, \citet{Lee_Hastie_2015} considered a mixed graphical model using node-conditionals to estimate the parameters in a graph that included both continuous and discrete nodes. \citet{chen2015selection} further generalized this type of model, allowing the conditional distribution of each node to belong to the exponential family. In their approach, some restrictions on the parameter space are required for the existence of a well-defined joint density. The authors claims that under truncation, it is possible to circumvent some of these restrictions, and estimate the structure of the graph even in the absence of a valid density. 
%An algorithm under this strategy was proposed by \citet{laman2021parallel}. This idea was originally implemented by \citet{yang2013poisson}, who proposed a truncation framework for the case of Poisson Graphical Models. 
However, truncation reduces the ability to generalize results and the estimates are sensitive to the truncation level.  Gaussian copula models have also been employed to model mixed types of data \citep{Hermes_van_2024, huge2021}. However, modeling multivariate counts with zero-inflated or multimodal marginals using a Gaussian copula is challenging. Contributions in the Bayesian framework for mixed types of data are relatively recent, and include the latent variable framework proposed by \citet{mohammadi2017bayesian} and a few approaches for mixed data that consider only continuous and categorical variables \citep{Bhadra_Rao_Baladandayuthapani_2018, Galimberti_Peluso_Castelletti_2024}.

Our purpose is to provide a more flexible mixed graphical modeling framework that can handle any type of data. For this, we adopt the conditionally specified modeling approach of \citet{chen2015selection} and circumvent the limitations described above by adopting the modeling strategy introduced by \citet{roy2020nonparametric} for count data, which allows estimation of both positive and negative interactions. We show how the resulting pairwise mixed graphical modeling framework can accommodate multiple types of data, including continuous, discrete, or categorical, and zero-inflated count data. We prove that the model maintains both global and local Markov properties. We employ the spike-and-slab prior of \citet{wang2015scaling} for the estimation of the graph and implement an MCMC algorithm for posterior inference. One of the advantages of our model formulation is that it captures the structure of the individual nodes, allowing us to impute missing data based on possible interactions that may exist between variables of different types. We exploit this aspect of the model and incorporate the imputation of missing data during the estimation process. We assess performances on four simulated scenarios with distinct dependence structures, that also include cases with data missing at random. We compare results with existing methods. We find that the proposed model matches or outperforms existing methods in cases with full data, while producing more consistent and robust performance in cases with missing data or when including zero-inflated count variables. Finally, we apply the proposed method to the real data collected from adolescents diagnosed with an eating disorder. Estimated graphs highlight the ability of the model to detect both positive and negative interactions between any type of data. We found differences in the associations estimated at intake and discharge, suggesting possible effects of the treatment on cognitive and behavioral measures in the adolescents.

The paper is structured as follows: In Section \ref{bmgm}, we describe the pairwise graphical model framework we propose. Sections \ref{prior_spc} and \ref{posterior_sam} describe the prior specification and the posterior sampling algorithm, respectively. In Section \ref{sec:simulations}, we provide results based on simulated data, and in Section \ref{sec:application}, we present the application to real data from adolescents diagnosed with an eating disorder. In Section \ref{sec:conclusion}, we provide some concluding remarks.

\section{Methods}
Suppose we have $p$ random variables, $\bm x=(x_1,\ldots,x_p)^T$, and let $G$ be the graph encoding the conditional independencies between the variables. We characterize the graph as $(V, E)$, with $V = \{1,...,p\}$ the vertex set, and $E \subset V \times V$ the set of edges of the graph. Thus, if two random variables, $x_s$ and $x_t$, with $(s,t) \in V \times V$, are conditionally independent given all the other variables, then $(s,t) \notin E$.

\subsection{Bayesian Mixed Graphical Model}
\label{bmgm} 
We propose a Bayesian pairwise graphical model that estimates conditional independencies between mixed types of data. This setting, known as Mixed Graphical Model, implies that the distributions of each node, conditioned upon the others, can belong to different family forms. Here, we consider multiple types, including continuous, discrete, or categorical, and zero-inflated count data. Our model formulation also allows for data missing at random. 

Let us start by introducing the generic pairwise graphical model representation, as described by \citet{wainwright2006high, chen2015selection} and \citet{roy2020nonparametric}. Thus, we write the joint density of $x$ as

\begin{equation}
    p(\bm x) \propto \exp\left\{\sum_{s=1}^p f_s(x_s) + \frac{1}{2}\sum_{s=1}^p \sum_{t\neq s} f_{ts}(x_s, x_t)\right\}, 
\label{eq:joint_org}
\end{equation}

\noindent where $f_s(x_s)$ is called the node-potential function and $f_{ts}(x_s, x_t)$ the edge-potential function with $(s, t) \in V \times V$. The edge-potential function satisfies the condition that if $(s, t) \notin E$, then $f_{st} = f_{ts} = 0$. The node-potential function can be set to target specific univariate marginal densities. For instance, in the case of the Poisson graphical 
model, it is given by $f_s(x_s) = \theta_s x_s - \log(x_s!)$ \citep{yang2013poisson}. 

The edge-potential function in \eqref{eq:joint_org} expresses the interactions between random variables. One common choice in the pairwise graphical modeling framework is to assume $f_{ts}(x_s, x_t) = \beta_{st} x_s x_t = \beta_{ts} x_s x_t$ \citep{wainwright2006high, hastie2015statistical, chen2015selection}. However, in the case of the Poisson graphical model, this edge potential function imposes some restrictions to the parameter space, allowing only negative values for the edge-weights $\beta_{st}$, and thus capturing only negative relationships between variables \citep{chen2015selection}.
To circumvent this problem, \citet{roy2020nonparametric} proposed the edge-potential function $f_{ts}(x_s, x_t) = \beta_{ts}F(x_t)F(x_s)$, with $F$ a monotone increasing bounded function with support $[0,\infty)$.  
In particular, the authors considered $F(x) = \tan^{-1}(x)^\lambda$ for some positive $\lambda \in \mathbb{R}^{+}$ to define a flexible class of monotone increasing bounded functions, and where $\lambda$ provides additional flexibility on the range of $F$. Thus, assuming this edge-potential function, the joint density in (\ref{eq:joint_org}) can be rewritten as
\begin{equation}
    p(\bm x) \propto \exp\left\{\sum_{s=1}^p f_s(x_s) + \frac{1}{2}\sum_{s=1}^p \sum_{t\neq s} \beta_{ts}F(x_t)F(x_s)\right\},
    \label{eq:joint}
\end{equation}

\noindent which has a finite normalizing constant, and allows unrestricted support for all the parameters, i.e., both positive and negative dependencies, in particular for discrete random variables.

For mixed graphical models, \citet{chen2015selection} suggested to utilize the framework of conditionally specified models, by first specifying the conditional densities $p(x_s | x_{-s})$ with ${\bm x}_{-s} = (x_1, ..., x_{s-1}, x_{s+1}, ..., x_p)$ and then combining the $p$ dimensional distributions to learn the parameters of the entire graph $G$. 
As shown by the authors, however, under the joint density in (\ref{eq:joint_org}), with edge potential function $f_{ts}(x_s, x_t) = \beta_{st} x_s x_t$, the conditionally specified model with node-conditional distributions corresponds to a well-defined density only in a few particular cases. This limits the ability to estimate the edge-potential parameters. For instance, under this proposal, it is not possible to determine edges between Gaussian and Poisson nodes, or Gaussian and Exponential nodes. Additionally, it only allows the estimation of negative interactions between two pairs of Poisson nodes or between Poisson and Exponential nodes \citep{chen2015selection}.  

In our proposed construction, we adopt the conditionally specified modeling framework of \citet{chen2015selection} but circumvent the limitations described above by adopting the edge-potential function $F(x_s) = \tan^{-1}(x_s)^\lambda$, if $x_s$ is a count or zero-inflated count variable, and $F(x_s) = x_s$ if $x_s$ is continuous or categorical, and therefore considering conditional densities of the form
\begin{equation}
\label{eq:conditional}
    p(x_s | \bm x_{-s}) = \exp\left\{f_s(x_s) + \sum_{t\neq s} \beta_{ts}F(x_t)F(x_s) - A\right\},
\end{equation}
where $\bm x_{-s} = (x_1, ..., x_{s-1}, x_{s+1}, ..., x_p)$, for $s \in V$, and with $A$ the normalizing constant. The following proposition ensures that the conditionally specified model with node-wise specification of the distributions given in \eqref{eq:conditional} corresponds to a well-defined joint density.  

\begin{proposition}
    Let $\bm x = (x_1, x_2,...,x_p)^T$ be a random vector, where for each $s$ the conditional density of $x_s$ given $\bm x_{-s}$ is of the form as in (\ref{eq:conditional}). If $\beta_{ts} = \beta_{st}$, and $F(0) = 0$, then the conditional densities are strongly compatible.  %, as defined by \citet{chen2015selection}. 
    Additionally, any function $g$ that is capable of generating the conditional densities is of the form

    \begin{equation}  
    g(\bm x) \propto \exp\left\{\sum_{s=1}^p f_s(x_s) + \frac{1}{2}\sum_{s=1}^p \sum_{t\neq s} \beta_{ts}F(x_t)F(x_s)\right\}.
    \end{equation}
\end{proposition}
\noindent Following \citet{chen2015selection}, we say that two conditional densities are compatible if there exists a function $g$ that is capable of generating both conditional densities. When $g$ is a density, the conditional densities are said to be strongly compatible. See the Appendix for the proof of the Proposition, following \citet{chen2015selection} and \citet{besag1974spatial}.
%The proof resembles the one presented in \citet{chen2015selection}, which follows the proof by \citet{besag1974spatial}, and can be found in the Appendix. 
This proposition ensures that the joint distribution $p(x)$ given in \eqref{eq:joint} is a valid density that is capable of generating the conditional densities in \eqref{eq:conditional}.

Furthermore, an important property of the proposed model is that if there exist two variables that are conditionally independent given the others, this is equivalent to the pair of interaction parameters being zero between the two corresponding nodes in the graph $G$. This is remarked in the following proposition. 
\begin{proposition}
    Consider that the joint density of $\bm x = (x_1, x_2 ,..., x_p)$ is given as in (\ref{eq:joint}). If $\beta_{ts} = \beta_{st}  = 0$, then $x_t$ and $x_s$ are conditionally independent given the other variables.
\end{proposition}
\noindent The proof is presented in the Appendix. This proposition implies that $\beta_{st} = \beta_{ts} = 0$ corresponds to a missing edge between nodes $s$ and $t$ in the graph $G$, which characterizes $p(x)$ as a pairwise graphical model. Also, the considered joint distribution (\ref{eq:joint}) satisfies the Hammersley-Clifford theorem. This theorem states that a probability distribution with a strictly positive density will exhibit a Markov property with respect to an undirected graph $G$ if and only if its density can be factorized over the cliques of the graph \citep{clifford1971a}. Since the pairwise graphical model considered can be represented as a product of mass functions of the cliques of graph $G$, this factorization guarantees that the distribution maintains both global and local Markov properties.

The proposed framework can accommodate multiple types of data, including continuous, discrete, or categorical, and zero-inflated count data. Next, we define the node potential functions $f_s(x_s)$ depending on the type of $x_s$: 
%\textcolor{red}{AG: I am not sure, but I was wandering: the node potential function is a function and not a density. Is it ok to use the conditioning symbol? maybe ; instead ?}

\begin{itemize}
    \item Count Data: $f_s(x_s;\bm \theta_s) = \nu_s(\log(\mu_s) x_s - \log(x_s!))$.
    \item Zero-Inflated counts: $f_s(x_s;\bm \theta_s) = \log(\pi_s\mathbbm{1}[x_s = 0] + (1-\pi_s)\exp(-\mu_s)\mu_s^{x_s}/x_s!)$.
    \item Continuous: $f_s(x_s;\bm \theta_s) = -\frac{1}{2}(\frac{x_s-\mu_s}{\sigma_s})^2 - \frac{1}{2}\log(\sigma_s^2)$. 
    \item Categorical: $f_s(x_s;\bm\theta_s) = \sum_{j=1}^k \mathbbm{1}[x_s = j] \log p_j$. 
\end{itemize}

\noindent These node potential functions capture the structure of different types of data. The node potential function for count data corresponds to the log of the Conway-Maxwell-Poisson distribution, which allows the modeling of data with different types of dispersion (\mbox{over-,} equi-, and under-dispersion) \citep{bensonfriel2021}. In this case, the parameters are determined by $\bm\theta_s = (\mu_s, \nu_s)$. Similarly, for zero-inflated count data, the node-potential targets the zero-inflated-Poisson distribution with $\bm\theta_s = (\pi_s, \mu_s)$. Additionally, $\bm\theta_s = (\mu_s, \sigma_s)$ allows to model continuous data and $\bm\theta_s = (p_1, p_2, ..., p_k)$ to model categorical variables.

\subsection{Prior Specification}
\label{prior_spc}
We assign standard independent priors on the node-potential parameters. Specifically, according to the node type, $x_s$, for $s \in V$, we consider the following priors:
\begin{itemize}
    \item Discrete counts: we assume $\mu_s \sim Gamma(u_0^s, v_0^s)$ and $\nu_s \sim Gamma(c_0^s, d_0^s)$.
    \item Zero-inflated counts: we let $\mu_s \sim Gamma(u_0^s, v_0^s)$, and $\pi_s \sim Beta(p_0^s, q_0^s)$.
    \item Continuous:  we let $\mu_s \sim N(\mu_0^s, {\sigma^2_0}^s)$, and $\sigma_s^2 \sim IG(a_0^s, b_0^s)$. 
    \item Categorical:  we let $p_1,p_2,...,p_k \sim Dirichlet(\alpha_1, \alpha_2, ..., \alpha_k)$. 
    \end{itemize}
In all simulations and applications of this paper, we allow non-informative priors, see Section \ref{sec:simsettings}. 
Following the strategy adopted by \citet{roy2020nonparametric} for  $F(\bm x) = (F(x_1), F(x_2), \ldots ,F(x_p))$,
we introduce the precision matrix $\Omega$, where $\Omega_{st} = \Omega_{ts} = \beta_{st}$, with $s < t$ and $\Omega_{ss} = (Var(F( \bm x))^{-1})_{ss}$.
%
%
%we assume that the standardized $F(x) = (F(x_1), F(x_2), \ldots ,F(x_p))$ follows a Gaussian 
%\textcolor{blue}{AG: I am still puzzled about this gaussianity. Notice that Roy and Dunson state: \textit{To generate $\beta$, we consider a new likelihood that the standardized \ldots follows a multivariate Gaussian distribution}. This is subtle, but they are not assuming joint Gaussianity. Rather, they are considering a \textit{new} likelihood. If we were assuming joint gaussianity of $F(x)$, and that the interaction parameters of $g(x)$ were exactly the same of this joint Gaussian distribution, we could directly use the ordinary Gaussian graphical model inferential procedures on $F(x)$. Am I missing something here? I tried to find in our paper how this assumption enters into our model, but we are not actually presenting the likelihood function. To cut to the chase, a solution would be to say: Adopting the same strategy/setting/tactic as \citet{roy2020nonparametric}, we consider/treat the standardized $F(x) = (F(x_1), F(x_2), \ldots ,F(x_p))$ as a Gaussian \ldots } distribution %with precision matrix $\Omega$, where $\Omega_{st} = \Omega_{ts} = \beta_{st}$, with $s < t$ and $\Omega_{ss} = (Var(F( x))^{-1})_{ss}$. 
On this precision matrix, we impose the continuous spike-and-slab prior proposed by \citet{wang2015scaling}, given by the following hierarchical model 

\begin{gather}
    \label{eq:prior_spikeslab1}
    p( \Omega |  Z, v_0, v_1, \lambda) = \{C(Z, v_0, v_1, \lambda)\}^{-1} \prod_{i<j} N(\beta_{ij} | 0, v_{z_{ij}}^2)\prod_i\text{Exp}(\beta_{ii}|\frac{\lambda}{2})\mathbbm{1}_{\{\Omega \in M+\}} \\
     \label{eq:prior_spikeslab2}
    p( Z|\Lambda) = \{C(\Lambda)\}^{-1}C(Z,v_0,v_1,\lambda)\prod_{i<j} \left \{\pi^{z_{ij}}(1-\pi)^{1 - z_{ij}} \right\},
\end{gather}

\noindent where $v_{z_{ij}} = v_0$ or $v_1$ if $z_{ij} = 0$ or $1$, respectively. The term $C(\Lambda)$, with $\Lambda = \{v_0, v_1,\pi,\lambda\}$, represents the normalizing constant that ensures integration over the space of positive definite matrices $M+$. The binary variables $Z = (z_{ij})_{i<j} = 1$ or $0$ indicate the presence or absence of an edge between nodes $i$ and $j$. These indicators are assigned independent Bernoulli priors, with equal inclusion probability $\pi$. In prior \eqref{eq:prior_spikeslab1} $v_0$ should be selected to ensure that if the data favors $z_{ij} = 0$ over $z_{ij} = 1$, then $\beta_{ij}$ is sufficiently small to be effectively replaced by zero. Likewise, $v_1$ should be chosen such that if the data indicates a preference for $z_{ij} = 1$ over $z_{ij} = 0$, then $\beta_{ij}$ can be accurately estimated to be significantly different from zero. 

\subsection{Posterior Sampling}
\label{posterior_sam}

For posterior inference, we implement an MCMC algorithm.  Furthermore, in designing our sampler, we accommodate data missing at random. We leverage closed-form expressions for the normalizing constant in the continuous and categorical cases, and employ numerical approximations for count and zero-inflated count data. Specifically, we note that the conditional likelihood at the \textit{s-th} node (\ref{eq:conditional}) can be expressed as 

\begin{equation}
    p(x_s | \bm x_{-s}) = \frac{q(x_s|\bm x_{-s}, \theta_s, {\bm \beta_s})}{A_q(\theta_s, {\bm \beta_s})},
\label{eq:intractable}
\end{equation}

\noindent where $q$ represents the unnormalized conditional likelihood and $A_q$ is the normalizing constant that depends on the parameters $\theta_s$ and $\bm \beta_s = \{\beta_{is}\}_{i \neq s}$. We next describe the updates of the individual parameters.

\subsubsection{Node-potential parameters}
\label{sec:node_potential_post}

We first update the node-specific parameters $\Theta = (\theta_1,\theta_2,\ldots,\theta_p)$ according to the type of each variable:
\begin{itemize}
    \item \textbf{Continuous variables.} For continuous nodes, the conditional distribution admits a closed-form expression for the normalizing constant:

    \begin{equation*}
    A_q(\theta_s,\beta_s) = \int_{-\infty}^{\infty} \exp\left(-\frac{1}{2\sigma_s^2}(x_s - \mu_s)^2 + x_s\sum_{t\neq s}\beta_{st}F(x_t)\right)dx_s
    \end{equation*}
    
    \noindent which is equal to $\sqrt{2\pi\sigma_s^2}\exp\left(C\mu_s + \frac{C^2\sigma_s^2}{2}\right)$, where $C = \sum_{t\neq s} \beta_{st} F(x_t)$. Then, a proposal $\theta_s^* = (\mu_s^*, \sigma_s^{2*})$ is accepted with probability: %\textcolor{purple}{This sentence is repeated twice. Please notice the different square in the sigmas}
    \begin{equation}
    \alpha(\theta_s, \theta_s^*) = \min\left\{1, 
    \frac{
    \prod_{i=1}^n p(x_{is} \mid \bm{x}_{i,-s}, \theta_s^*, \bm{\beta}_s) \cdot p(\theta_s^*)}{
    \prod_{i=1}^n p(x_{is} \mid \bm{x}_{i,-s}, \theta_s, \bm{\beta}_s) \cdot p(\theta_s)}\right\}.
    \end{equation}

    \item \textbf{Discrete count variables.} For count variables, the normalizing constant is not available in closed form. We adopt the approach proposed by \citet{roy2020nonparametric} and truncate the normalizing constant at a sufficiently large value $B = 100$, 
    \begin{equation}
    A_q(\theta_s, \bm{\beta}_s) \approx \sum_{x=0}^{B} \exp\left\{ \nu_s \log(\mu_s)x - \nu_s \log(x!) - \sum_{t \neq s} \beta_{st} F(x_{t})F(x) \right\}.
    \end{equation}
    A proposal \(\theta_s^*\) is accepted with probability:
    \begin{equation}
    \alpha(\theta_s, \theta_s^*) = \min\left\{1, 
    \frac{
    \prod_{i=1}^n q(x_{is} \mid \bm{x}_{i,-s}, \theta_s^*, \bm{\beta}_s) \cdot p(\theta_s^*) \cdot A_q(\theta_s, \bm{\beta}_s)
    }{
    \prod_{i=1}^n q(x_{is} \mid \bm{x}_{i,-s}, \theta_s, \bm{\beta}_s) \cdot p(\theta_s) \cdot A_q(\theta_s^*, \bm{\beta}_s)
    }\right\}.
    \end{equation}

    \item \textbf{Zero-inflated count variables.} The normalizing constant is approximated in the same way as for count variables. Proposals $\theta_s = (\pi_s^*, \lambda_s^*)$ are accepted using the same Metropolis-Hastings formulation.

    \item \textbf{Categorical variables.} For nodes with $K$ categories, the normalized conditional likelihood is given by:
    \begin{equation}
    p(x_{is} = k \mid \bm{x}_{i,-s}) = \frac{ \theta_{sk} \exp\left\{ -\sum_{t \neq s} \beta_{st}^{(k)} F(x_{it}) \right\} }{ \sum_{j = 1}^{K} \theta_{sj} \exp\left\{ -\sum_{t \neq s} \beta_{st}^{(j)} F(x_{it}) \right\} }.
    \end{equation}
    The normalizing constant is available in closed form. A proposed $\theta_s^*$ is accepted with probability:
    \begin{equation}
    \alpha(\theta_s, \theta_s^*) = \min\left\{1, 
    \frac{
    \prod_{i=1}^n p(x_{is} \mid \bm{x}_{i,-s}, \theta_s^*, \bm{\beta}_s) \cdot p(\theta_s^*)
    }{
    \prod_{i=1}^n p(x_{is} \mid \bm{x}_{i,-s}, \theta_s, \bm{\beta}_s) \cdot p(\theta_s)
    }\right\}.
    \end{equation}
\end{itemize}

\subsubsection{Edge-potential parameters}

To update the edge-specific weight parameters, we use the Gibbs sampler proposed by  \citet{wang2015scaling}, which updates each column of the precision matrix $\Omega$ successively. Since the elements $\Omega_{st} = \Omega_{ts} = \beta_{st}$, and the diagonal entries do not change over iterations, updating the \textit{l-th} column of $\Omega$, ${\Omega_l}$, corresponds to update the edge-weight parameters $\bm \beta_l = \{\beta_{il}\}_{i \neq l}$. We let $\Omega_{-l-l}$ denote the submatrix of $\Omega$ removing the \textit{l-th} row and column, and consider $G$ the gram matrix of $F(X)$, $G = (F(x) - \bar{F}(x))^T(F(x) - \bar{F}(x))$. Thus, we update each column $\Omega_l$, $1 \leq l \leq p$, of $\Omega$ successively as follows:

\begin{itemize}
    \item[i.] Generate an update ${\Omega_l}^*$ for $\Omega_l$. A candidate is generated from a MVN$(-C G_{l,-l}, C)$, where $C = (G_{ll}\Omega_{-l-l}^{-1} + D_l^{-1})^{-1}$, with $D_l$ the prior variance corresponding to $\Omega_{l,-l}$. 
    \item[ii.] Compute the acceptance probability of ${\Omega_l}^* = {\bm \beta_l}^* = \{{\beta_{il}}^*\}_{i \neq l}$ as:
    \begin{equation*}
        \min \left\{1, \frac{\prod_{i = 1}^{n} q(x_{il} | \bm x_{i-l},  \theta_l,  \bm \beta_l^*)A_q(\theta_l,\beta_l)\cdot p(\bm \beta_{l}^*) }
        {\prod_{i = 1}^{n} {q(x_{il} | \bm x_{i-l},  \theta_l,  \bm \beta_l)}A_q(\theta_l,\beta_l^*)\cdot p( \bm \beta_l)}  \right\}
    \end{equation*}
    The form of $A_q$ depends on the type of variable at node $l$, with closed-form expressions for continuous and categorical variables, and numerical approximations otherwise as defined in Section \ref{sec:node_potential_post}.

    \item[iv.] Sample the indicators $z_{ij}$. As pointed out by \citet{wang2015scaling}, priors \eqref{eq:prior_spikeslab1} and \eqref{eq:prior_spikeslab2} imply that the $z_{ij}$'s are distributed as independent Bernoulli distributions, with probability
    \begin{equation}
        Pr(z_{ij} = 1 |  \beta,  x) = \frac{N(\beta_{ij}| 0, v_1^2)\pi}{N(\beta_{ij}| 0, v_1^2)\pi + N(\beta_{ij}| 0, v_0^2)(1-\pi)}.
    \end{equation}
\end{itemize}

\subsubsection{Missing data}
Our application dataset in Section \ref{sec:application} is characterized by missing data. Therefore, we allow for missingness in our sampling procedure by considering the following multiple imputation model. We first assume that the missing data pattern is missing at random (MAR), which refers to the condition when the missingness probability depends on observed information. Under this assumption, it is said that the missing data mechanism is \textit{ignorable} \citep{little2019statistical}. We then follow the data augmentation algorithm described by \citet{tanner1987calculation} and \citet{van2018flexible}, which consists of the following steps:
\begin{itemize}
\item Let $\bm x = (x_{mis}, x_{obs})$, with $x_{mis}$ representing the incomplete data and $x_{obs}$ the observed data. At iteration $t$, given the sampled $\beta^t$ and $\theta^t = (\theta_1^t,...,\theta_p^t)$, we draw a value of the missing data from the conditional predictive distribution of 
\begin{eqnarray}
x_{mis} \sim p(x_{mis} | x_{obs}, \theta^t, \beta^t).
\end{eqnarray} 

We implement this step by using a Gibbs sampler through the conditional likelihoods defined in (\ref{eq:conditional}). We use $m = 10$ complete datasets simulated at each step that are later combined by taking their element-wise mean to produce a single overall inference. 
\item Then, conditioning on $x_{mis}$, we draw new values for $\theta$ and $\beta$ through the conditional likelihoods as described previously.
\end{itemize}

\subsection{Posterior inference}
\label{sec:post_inf}
To perform edge selection, we calculate the posterior expected false discovery rate (FDR) following an approach similar to what proposed by \citet{newton2004detecting}. Specifically, we first calculate the posterior probability of inclusion (PPI) as the proportion of MCMC iterations, discarding burn-in, in which the edges were included in the graph. Then, we employ the posterior expected FDR to determine the probability cutoff for the posterior probability of inclusion. The posterior expected FDR is given by 

\begin{equation}
    E[FDR_c \mid data] = \frac{\sum_{i<j}(1-p_{ij})\mathbbm{1}(p_{ij} > c)}{\sum_{i<j}\mathbbm{1}(p_{ij} > c)},
\end{equation}

\noindent with $p_{ij}$ the PPI of the edge between nodes $i$ and $j$, and where the cutoff $c$ is chosen such that $\min\{c : E[FDR_c \mid data] \leq 0.05\}.$

\section{Simulations}
\label{sec:simulations}
We test performances of our proposed model by simulating mixed data with different dependence structures, performing a comparison with existing methods.

\subsection{Parameter Settings}
\label{sec:simsettings}
In all simulations presented below, we fit our proposed model, which we call BMGM, using non-informative priors on the node-potential parameters, as specified in Section \ref{prior_spc}. Specifically, we set $\mu_0^s = 0$, ${\sigma^2_0}^s = 100$ and $a_0^s = b_0^s = 0.001$ for continuous nodes, and $u_0^s = v_0^s = 0.001$ for the location parameter $\mu_s$ in the case of discrete count and zero-inflated count nodes. For the dispersion parameter $\nu_s$ in the case of discrete count nodes, we considered $c_0^s = d_0^s = 0.001$, and, similarly, we let $ p_0^s = q_0^s = 0.001$ for the zero-inflation parameter $\pi_s$. For categorical nodes, we let $\alpha_j = \frac{1}{k}, 1 \leq j \leq k$. Furthermore, we used standard values $v_0 = 0.05$, $v_1 = 1$, $\lambda = 1$, and $\pi = \frac{2}{p-1}$, with $p$ the number of nodes, for the prior \eqref{eq:prior_spikeslab1} on the precision matrix $\Omega$, as suggested by \citet{wang2015scaling}. When running MCMC chains, the initial values of the non-diagonal elements of $\Omega$ were chosen to be 0, and we allowed 10,000 iterations of burn-in, followed by another 10,000 iterations for analysis. The algorithm was tuned such that the acceptance rates were between $20\%$ and $60\%$. For edge-weight parameters, we applied a rescaling step following \citet{roy2020nonparametric}, whereas for node-potential parameters, proposal variances were adaptively tuned based on empirical covariances following standard MCMC practice.

For comparison with our method, we used the pairwise mixed graphical model proposed by \citet{mgm2020}, available in the R package \textit{mgm}, version 1.2.14, the non-paranormal graphical lasso, using the R package \textit{huge} \citep{huge2021}, version 1.3.5, and the Bayesian copula graphical model introduced by \citet{mohammadi2019bdgraph} contained in the R package \textit{BDgraph}, version 2.72. For all methods, we considered the default parameters.

\subsection{Simulation 1}
\label{sec:random_graph}

In this scenario, we simulated multivariate mixed data using a Gibbs sampler over the conditional densities specified in (\ref{eq:conditional}). We considered two cases, $p = 12$ and $p = 20$, each comprising $p/2$ continuous nodes and $p/2$ discrete nodes. We also tried different sample sizes, i.e., $n=200; 500$. The node potential parameters for the continuous nodes were set to $\mu_s$ = 0, with $\sigma_s$ randomly sampled from the set $\{1, 2, 4\}$. Similarly, for count nodes we let $\mu_s = 1$ and $\nu_s$ randomly chosen from $\{0.5, 1, 2\}$ to account for over-, equi-, and under-dispersed data. For the edge potential parameters between nodes $s$ and $t$, we let 
\begin{gather*}
    \beta_{ts} = \beta_{st} = z_{st}y_{st}r_{st},\\
    z_{st} \sim \text{Bernoulli}\left(\pi = \frac{2}{p-1}\right), \\
    r_{st} \sim \mathcal{U}(0.2, 0.6),\\
    Pr(y_{st} = 1) = Pr(y_{st} = -1) = 0.5.
\end{gather*}
This formulation allows for both positive and negative interactions between nodes.  

\begin{figure}
\centering
\includegraphics[width=1\linewidth]{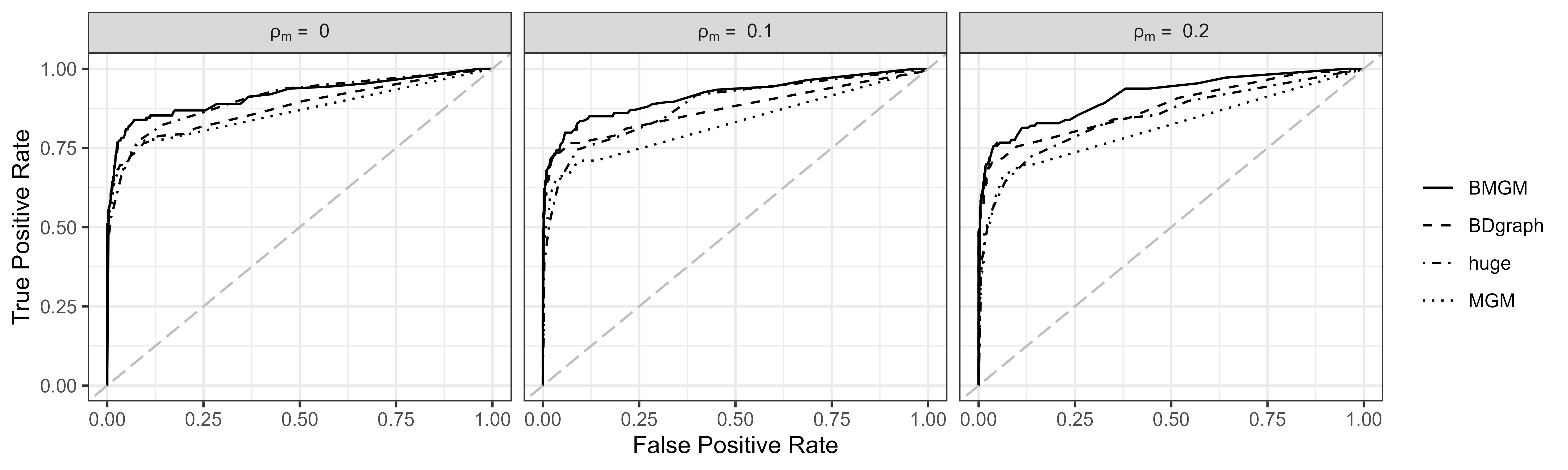}
\includegraphics[width=1\linewidth]{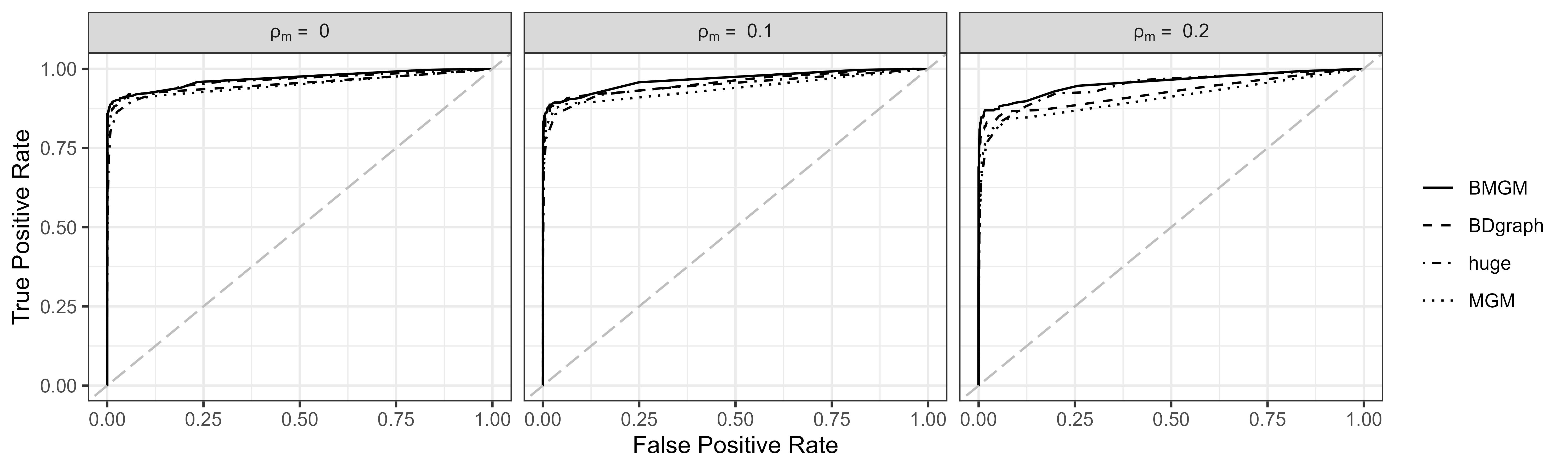}
\includegraphics[width=1\linewidth]{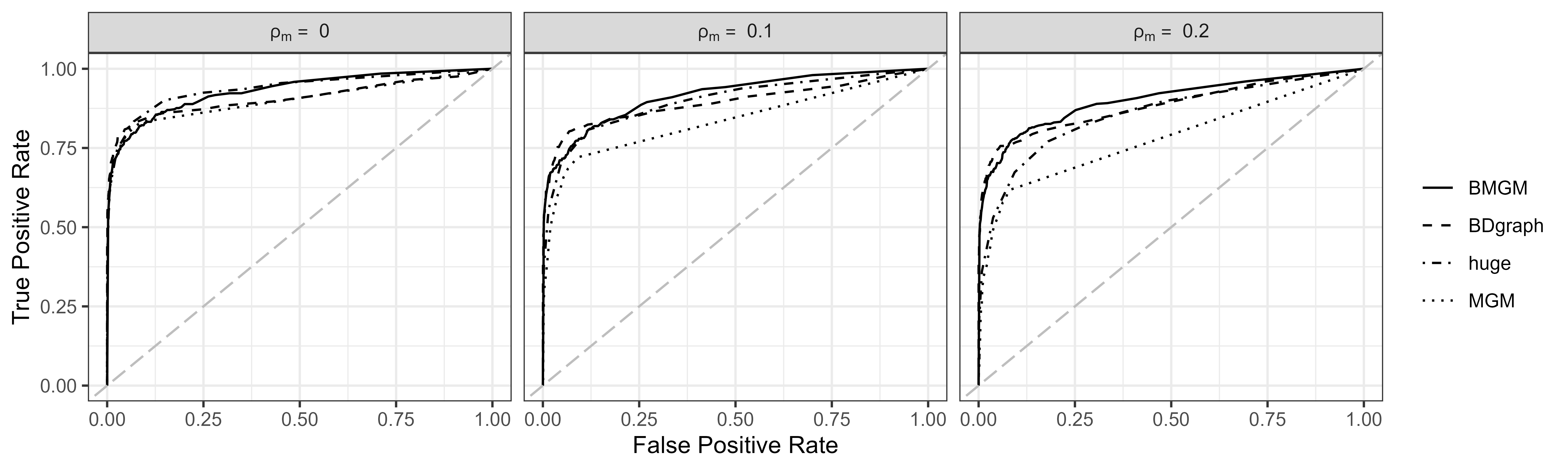}
\includegraphics[width=1\linewidth]{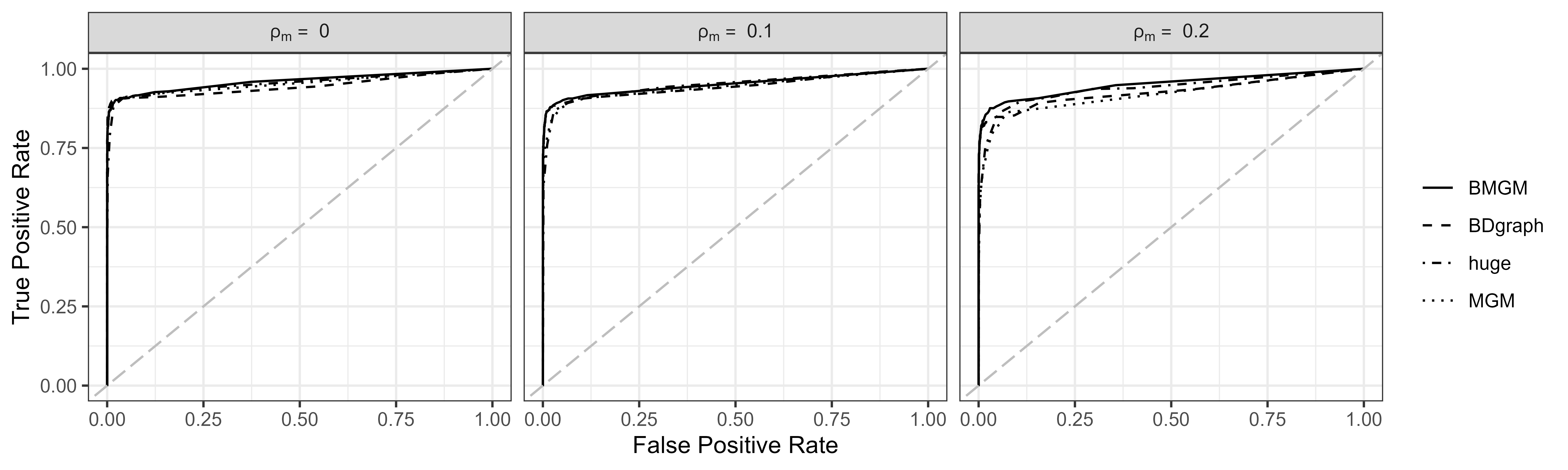}
\caption{Simulation 1: ROC curves for $p = 12$ and $n = 200; 500$ (top two rows) and for $p = 20$ and $n = 200; 500$ (bottom two rows), averaged across 30 replicates. The first column corresponds to the full data case $\rho_{m}$ = 0, the second and third columns to cases $\rho_{m} = 0.1; 0.2$, that have $10\%$ and $20\%$ of the observations with missing values in the chosen nodes, respectively. Black lines $(\sampleline{})$ indicate ROC curves of the proposed method \textit{BMGM}, $(\sampleline{dashed})$ lines to \textit{BDgraph} \citep{mohammadi2019bdgraph}, $(\sampleline{dash pattern=on .3em off .2em on .05em off .2em})$ lines to the \textit{huge} method \citep{huge2021}, and $(\sampleline{dotted})$ lines to \textit{MGM} \citep{mgm2020}. } 
\label{fig:sim_random_roc}
\end{figure}

As previously highlighted, one of the advantages of our model formulation is that it captures the structure of the individual nodes, allowing us to impute missing data based on possible interactions that may exist between variables of different types. Therefore, we also considered cases with data missing at random to test how well the model recovers the observed connections without complete data. More specifically, let $V_{int} \subseteq V $ denote the subset of nodes in $V$ that have interactions (i.e., nodes that have at least one edge in $E$), we randomly induced missing data using a censoring mechanism \citep{santos2019missing} in half of the nodes in $V_{int}$, which corresponds to around 5 or 6 count and continuous variables, respectively. We considered three scenarios, $\rho_{m} = 0$ that corresponds to full data, $\rho_{m} = 0.1$, and $\rho_{m} = 0.2$, where 10\% and 20\% of the observations in the chosen nodes have missing data, respectively. When applying the competing methods, we considered the data with complete cases (eliminating NAs) for MGM \citep{mgm2020} and Huge \citep{huge2021}, as those methods do not handle missing data, while we included NAs in BDgraph \citep{mohammadi2019bdgraph}.

\begin{table}
\centering
\begin{tabular}{@{}ccllll@{}}
\toprule
&&&$p=12$\\
\midrule
 $n$ & $\rho_{m}$ & BMGM  & BDgraph & huge  & MGM   \\ \midrule
\multirow{3}{*}{200} & 0     & \textbf{0.911} (0.09) & 0.880 (0.097) & 0.897 (0.08)  & 0.854 (0.1)\\
                     & 0.1   & \textbf{0.909} (0.097) & 0.872 (0.117) & 0.878 (0.095) & 0.825 (0.114) \\
                     & 0.2   & \textbf{0.912} (0.09)  & 0.880 (0.1)   & 0.838 (0.12) & 0.815 (0.114)\\ \midrule 
\multirow{3}{*}{500} & 0     & \textbf{0.962} (0.06)& 0.958 (0.083)   & 0.954 (0.066) & 0.941 (0.083) \\ 
                     & 0.1   & \textbf{0.963} (0.063) & 0.942 (0.077)  & 0.932 (0.077)  & 0.931 (0.08) \\
                     & 0.2   & \textbf{0.949} (0.059)  & 0.931 (0.088) & 0.935 (0.055) & 0.893 (0.12) \\ 
\end{tabular}
\begin{tabular}{@{}ccllll@{}}
\toprule
&&&$p=20$\\
\midrule
 $n$ & $\rho_{m}$ & BMGM  & BDgraph & huge  & MGM   \\ \midrule
\multirow{3}{*}{200} & 0     & 0.932 (0.051) & 0.907 (0.083) & \textbf{0.933} (0.058) & 0.900 (0.074)\\
                     & 0.1   & \textbf{0.916} (0.04) & 0.894 (0.078) & 0.896 (0.057) & 0.831 (0.077) \\
                     & 0.2   & \textbf{0.903} (0.044)  & 0.887 (0.058)   & 0.856 (0.064) & 0.772 (0.066)\\ \midrule
\multirow{3}{*}{500} & 0     & \textbf{0.958} (0.059) & 0.947 (0.074)   & 0.953 (0.054) & 0.950 (0.055) \\ 
                     & 0.1   & \textbf{0.949} (0.065) & 0.948 (0.054)  & 0.945 (0.053)  & 0.942 (0.053) \\
                     & 0.2   & \textbf{0.949} (0.05)  & 0.928 (0.069) & 0.936 (0.055) & 0.917 (0.058) \\ \bottomrule
\end{tabular}
\caption{Simulation 1: AUC values (means and std in parentheses) for $p = 12; 20$ and $n=200; 500$, averaged across 20 replicates. The highest value in each scenario $(n, \rho_{m})$ is highlighted in bold. The full data case corresponds to $\rho_{m} = 0$, cases with $10\%$ and $20\%$ of the observations with missing values in the chosen nodes correspond to $\rho_{m} = 0.1; 0.2$, respectively.}
\label{tab:sim_random_auc}
\end{table}

Figure \ref{fig:sim_random_roc} shows the ROC curves calculated across 30 replicated datasets for the cases $p = 12$ and $p=20$, with sample sizes $n=200; 500$, and Table \ref{tab:sim_random_auc} reports the AUC values. We observe that our model generally outperforms the pairwise graphical model MGM across all scenarios. Also, while copula methods such as BDgraph and huge typically exhibit strong performance with complete data, they tend to falter in the presence of missing data. In contrast, our model demonstrates consistent and satisfactory performance across all scenarios, especially in the presence of missing data, where it attains the largest AUC values in comparison with the other methods. 

%{\color{red}@Anna, @Michele, we have moved the tables with MCC, sens and spec to the end of the paper but my inclination is not to present those results (only ROC and AUC) because we do not do well.}

\subsection{Simulation 2}
\label{sec:fixed_graph}

\begin{figure}
\centering
\includegraphics[scale = 0.51]{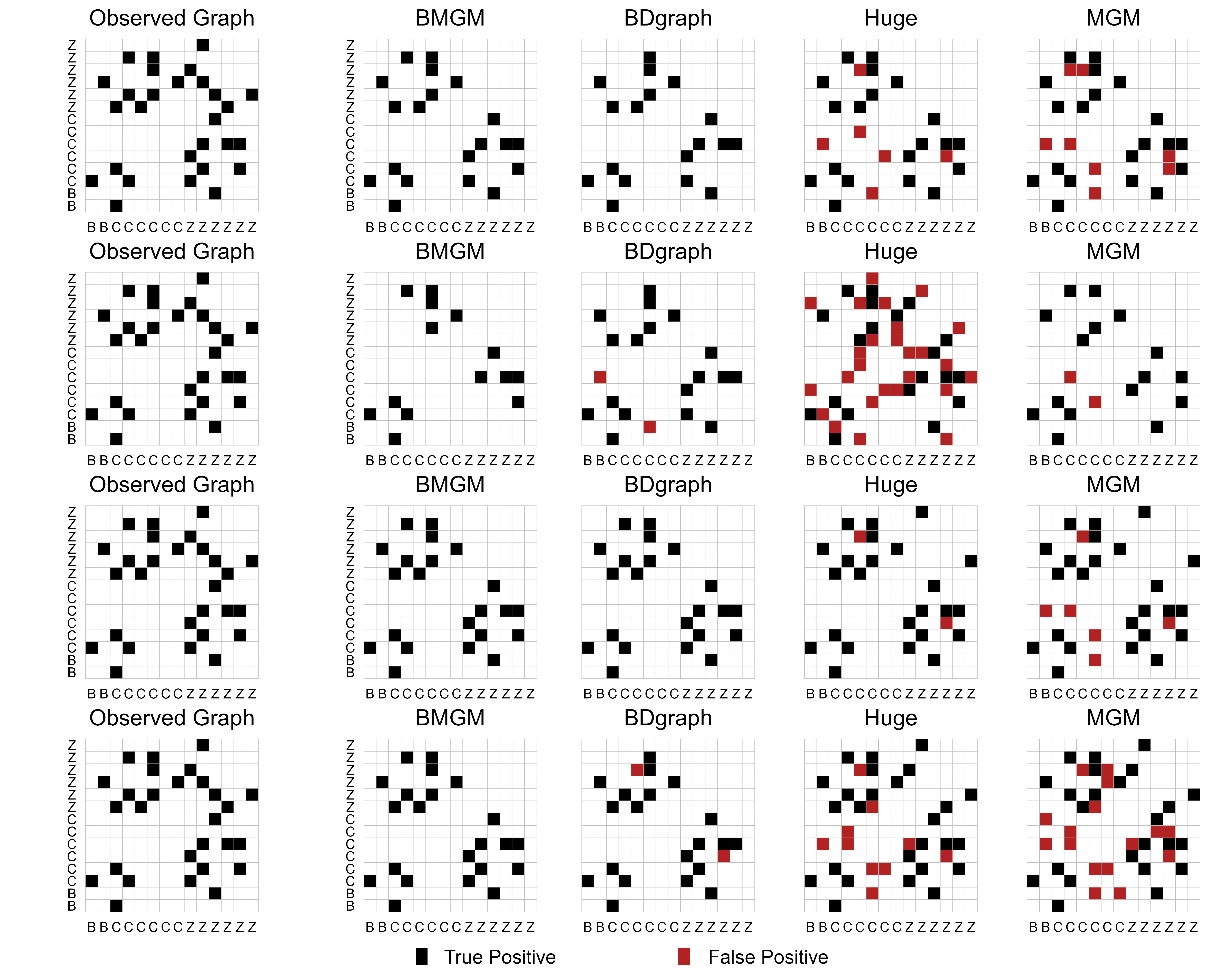}
    \caption{Simulation 2: Adjacency matrix of the true graph (first column) and the adjacency matrices of the estimated graphs by our method and competing approaches (columns 2-5), for one replicated dataset with $p = 14$, $n = 200$ and $\rho_{m} = 0; 0.1$ (top two rows) and $p = 14$, $n = 500$ and $\rho_m = 0; 0.1$ (bottom two rows). True positives are shown in black and false positives in red. The labels in the adjacency matrices indicate the type of variable: B = binary/categorical, C = continuous, and Z = zero-inflated count. Graph selection criteria were applied with the default options for each method in the R packages \textit{huge}, \textit{BDgraph}, and \textit{MGM}. For the BMGM method, refer to Section \ref{sec:post_inf}.} 
\label{fig:fixed_graph_est}
\end{figure}

For this scenario, we simulated data with the same dimensions and structure of the real data presented in Section \ref{sec:application}, that is $p = 14$, with 6 continuous, 6 zero-inflated counts, and 2 categorical variables. Unlike in the previous simulated scenario, where we randomly sampled the true graph, here we fixed positive and negative interactions between the variables as depicted in the "True graph" of Figure \ref{fig:fixed_graph_est}, see column 1. Similar to the previous simulated scenario, we considered the case of full data and the case where half of the nodes have $10\%$ missing values $(\rho_{m} = 0.1)$. Figure \ref{fig:fixed_graph_est} shows the estimated graphs by our method and the competing approaches, for one replicated dataset, for the $n=200; 500$ cases.  We can see that the proposed BMGM method is more robust to sample size and has fewer false positives than the other methods in both cases. 

\begin{figure}
    \centering
    \includegraphics[scale = 0.55]{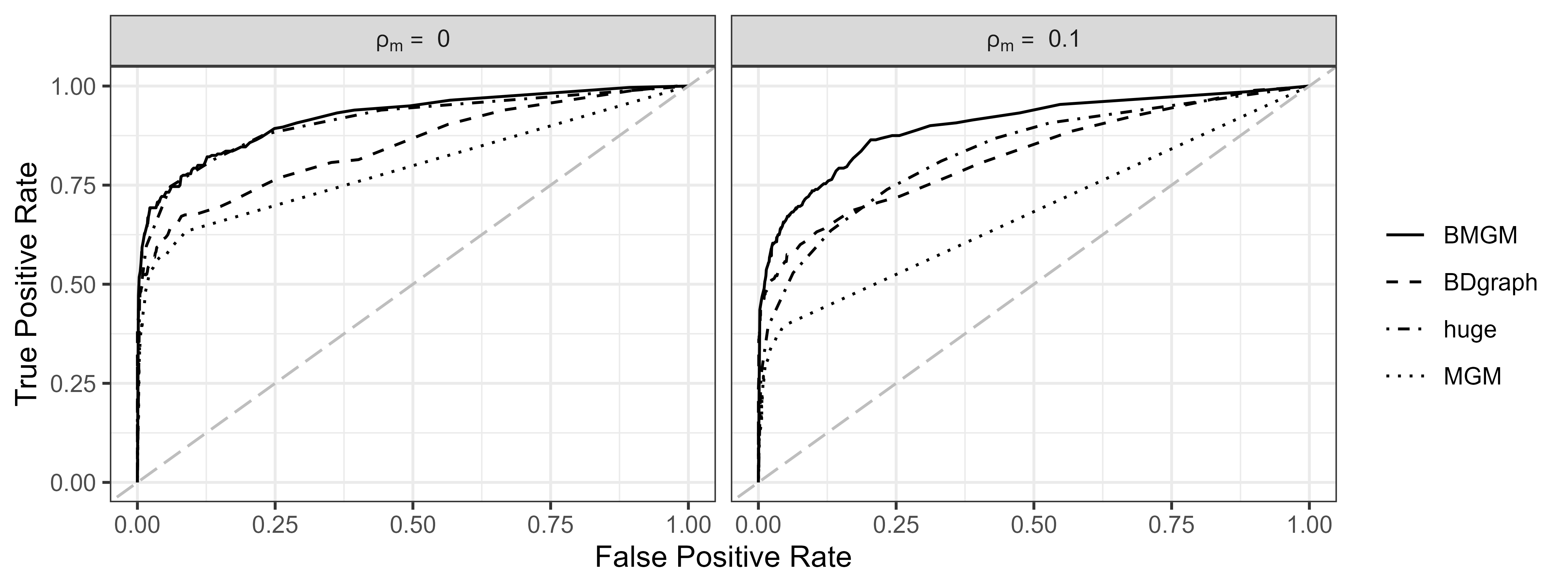}
    \includegraphics[scale = 0.55]{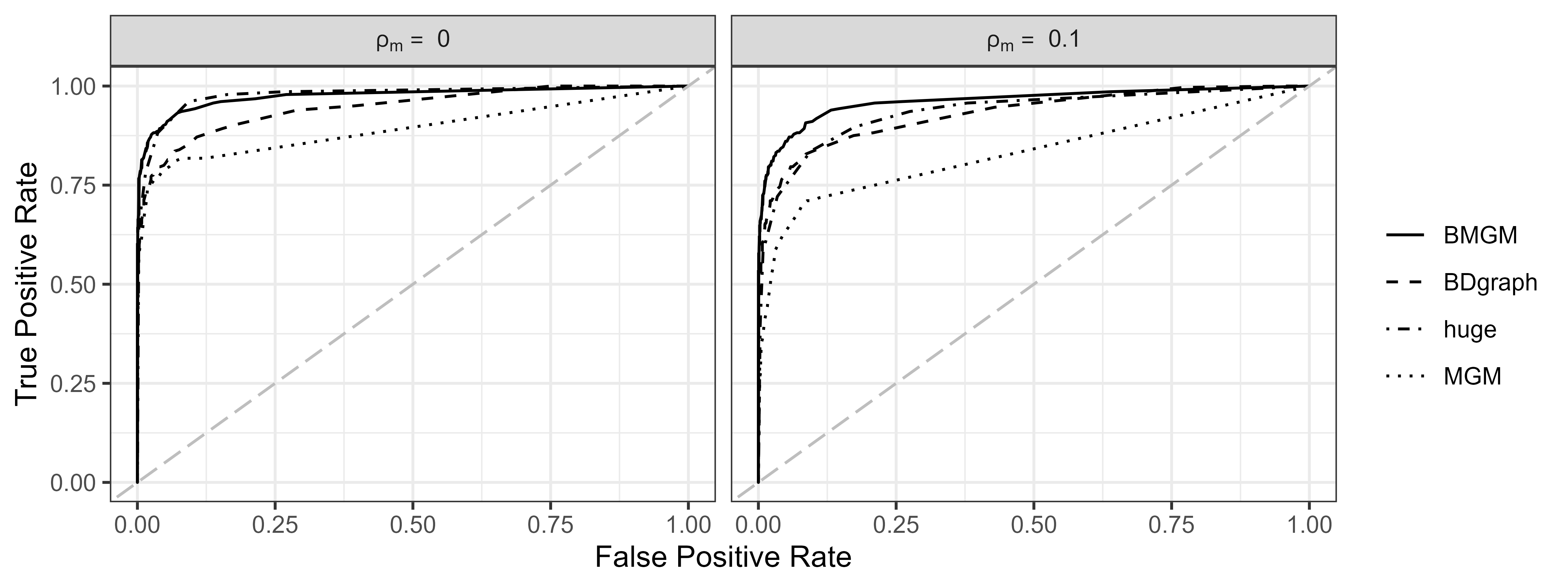}
    \caption{Simulation 2: ROC curves for $p = 14$ and $n = 200; 500$ (top and bottom row). The first column corresponds to the full data case $\rho_{m}$ = 0, and the second column to case $\rho_{m} = 0.1$ that corresponds to $10\%$ of the observations with missing values in the chosen nodes. The black lines $(\sampleline{})$ correspond to the ROC curves of the proposed method \textit{BMGM}, the $(\sampleline{dashed})$ lines to \textit{BDgraph} \citep{mohammadi2019bdgraph}, the $(\sampleline{dash pattern=on .3em off .2em on .05em off .2em})$ lines to \textit{huge} method \citep{huge2021}, and the $(\sampleline{dotted})$ lines to \textit{MGM} \citep{mgm2020}.}
    \label{fig:sim_fixed_roc}
\end{figure}

%{\color{blue} I wanted to plot the fixed graph as well as the proportion of correctly estimated edges. Now, with the result for one replicate, it is not what I expected... maybe we can choose to only present the ROC curves as in the Random Graph.} {\color{red}, Present results with the same figures/plots as for the previous simulation. Put the complete tables at the end of the paper.} We observe as well as the proportion of edges estimated across the replicates for each method in the case of $n = 200$.
Figure \ref{fig:sim_fixed_roc} shows the ROC curves calculated across 20 replicated datasets for sample sizes $n=200; 500$, and Table \ref{tab:sim_fixed_auc} reports the AUC values. From the ROC curves, it is evident that the proposed model outperforms the existing methods, including the case with full data. Also, when the sample size is increased, the performance of our model improves noticeably. This is likely to be caused by the inclusion of the zero-inflated count data, a challenge for copula models as noted by \citet{roy2020nonparametric}. Furthermore, our model attains the highest AUC values across all cases. Notably, the performance of the other methods declines substantially in the presence of missing data $(\rho_{m} = 0.1)$, whereas our model maintains adequate performance. These results indicate that our model is the most accurate at capturing the interactions between mixed data in this setting of variables while also demonstrating its robustness to missing data compared to competing methods.

\begin{table}[ht]
\centering
\begin{tabular}{@{}ccllll@{}}
\toprule
 $n$ & $\rho_{m}$ & BMGM & BDgraph & huge  & MGM   \\ \midrule
\multirow{2}{*}{200} & 0     & \textbf{0.916} (0.048) & 0.849 (0.055) & 0.905 (0.045) & 0.787 (0.056)\\
                     & 0.1   & \textbf{0.895} (0.04) & 0.828 (0.074) & 0.828 (0.053) & 0.677 (0.053) \\ \midrule
\multirow{2}{*}{500} & 0     & 0.97 (0.031) & 0.946 (0.034) & \textbf{0.973} (0.023) & 0.887 (0.037) \\ 
                     & 0.1   & \textbf{0.958} (0.035) & 0.93 (0.035) & 0.931 (0.036)  & 0.823 (0.046) \\ \bottomrule
\end{tabular}
\caption{Simulation 2: AUC values (means and std in parentheses) for $n=200; 500$, averaged across 20 replicates. The highest value in each scenario $(n, \rho_{m})$ is highlighted in bold. The full data case corresponds to $\rho_{m}$ = 0, the case with $10\%$ of the observations with missing values in the chosen nodes corresponds to $\rho_{m} = 0.1$.}
\label{tab:sim_fixed_auc}
\end{table}

%\subsection{Additional Simulations}

\subsection{Simulation 3}

To assess the robustness of our method under model misspecification, we conducted a third simulation study where the data were not generated from our proposed model. Instead, we simulated multivariate mixed data using the \textit{bdgraph.sim()} function from the BDgraph package of \cite{bdgraph2019}, which relies on a copula-based formulation and does not follow the conditional distribution structure of BMGM. The generated datasets comprised $p=10$ variables, including $4$ Poisson, $4$ Gaussian and $2$ categorical nodes with 2 levels each. The edge structure was randomly generated by the \textit{bdgraph.sim()} function. We considered two sample sizes, $n = 200$ and $n = 500$, and three scenarios: full data $(\rho_m = 0)$, and two levels of missing at random affecting $10\%$ and $20\%$ of observations in selected nodes $(\rho_m = 0.1, 0.2)$, as described in Section \ref{sec:random_graph}. 

\begin{figure}[ht]
\centering
\includegraphics[scale = 0.5]{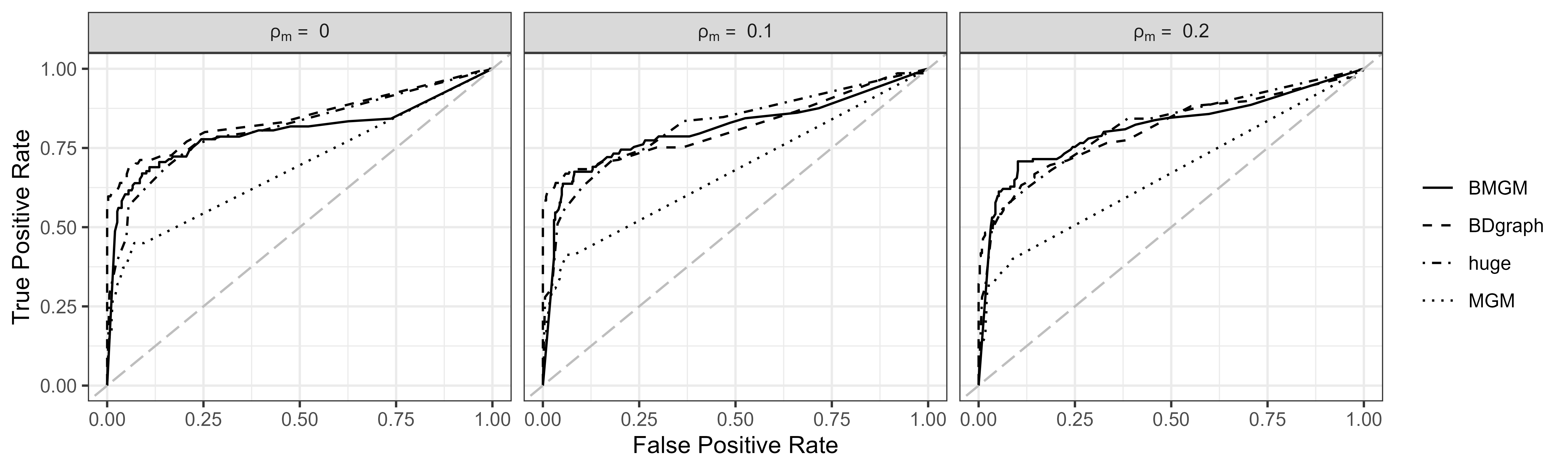}
\includegraphics[scale = 0.5]{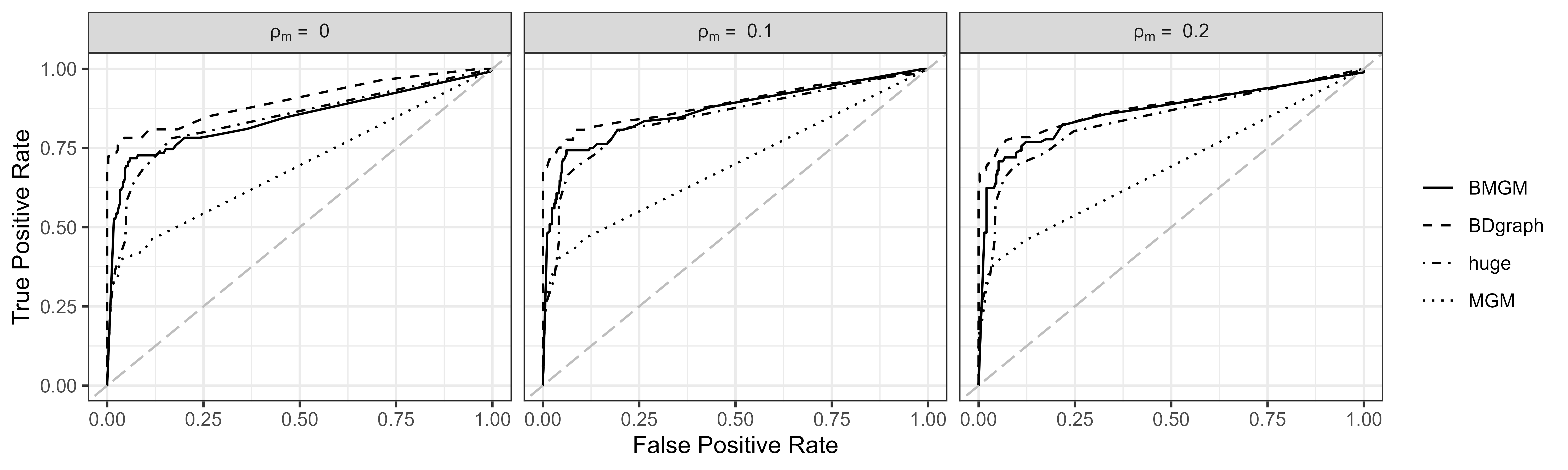}
\caption{Simulation 3: ROC curves for $p = 10$ and $n = 200$ (top row) and $n = 500$ (bottom row), averaged across 10 replicates. Data were generated from a BDgraph model. The first column corresponds to the full data case $\rho_{m} = 0$, while the second and third columns represent scenarios with $10\%$ and $20\%$ missingness in selected nodes, respectively ($\rho_{m} = 0.1, 0.2$). Black lines $(\sampleline{})$ indicate ROC curves of the proposed method \textit{BMGM}, $(\sampleline{dashed})$ lines correspond to \textit{BDgraph} \citep{mohammadi2019bdgraph}, $(\sampleline{dash pattern=on .3em off .2em on .05em off .2em})$ lines to the \textit{huge} method \citep{huge2021}, and $(\sampleline{dotted})$ lines to \textit{MGM} \citep{mgm2020}.
}

\label{fig:sim_bdgraph_roc}
\end{figure}

Figure \ref{fig:sim_bdgraph_roc} displays the ROC curves across $K = 10$ replicates for both sample sizes and level of missingness. We observe that BMGM achieves comparable or higher true positive rates than other approaches while maintaining a low false positive rate. In particular, for $n =200$, BMGM outperforms other methods with missing data, and for $n=500$ BMGM closely matches the performance of BDgraph,  despite the fact that the data were generated from BDgraph. To complement this, Table \ref{tab:sim_bdgraph_auc} shows the AUC values for each method. We find that BMGM performs on par with BDgraph and generally outperforms Huge and MGM across all settings. This simulation supports the robustness of BMGM to moderate levels of model misspecification, and underscore the competitive its performance and robustness under various simulation mechanisms.

\begin{table}[ht]
\centering
\begin{tabular}{@{}ccllll@{}}
\toprule
$n$ & $\rho_{m}$ & BMGM & BDgraph & huge  & MGM   \\ \midrule
\multirow{3}{*}{200} 
& 0     & 0.801 (0.125) & \textbf{0.846} (0.075) & 0.814 (0.12) & 0.687 (0.143) \\
& 0.1   & 0.813 (0.107) & \textbf{0.821} (0.082) & 0.817 (0.113) & 0.672 (0.134) \\
& 0.2   & \textbf{0.818} (0.108) & 0.808 (0.098) & 0.816 (0.116) & 0.664 (0.121) \\ \midrule
%& 0.3   & \textbf{0.820} (0.113) & 0.804 (0.085) & 0.812 (0.119) & 0.653 (0.133) \\ 

\multirow{3}{*}{500} 
& 0     & 0.843 (0.109) & \textbf{0.899} (0.072) & 0.838 (0.101) & 0.689 (0.129) \\
& 0.1   & 0.874 (0.098) & \textbf{0.890} (0.074) & 0.846 (0.098) & 0.694 (0.122) \\
& 0.2   & 0.870 (0.105) &  \textbf{0.879} (0.085) & 0.835 (0.1) & 0.685 (0.131) \\
%& 0.3   & \textbf{0.874} (0.101) & \textbf{0.879} (0.059) & 0.839 (0.091) & 0.673 (0.131) \\ 
\bottomrule
\end{tabular}
\caption{Simulation 3: AUC values (mean and SD in parentheses) for $n = 200; 500$, $p = 10$, and increasing missingness levels $\rho_{m}$. Data were generated from BDgraph. Values are averaged across $K=10$ replicates. The highest AUC value in each scenario $(n, \rho_{m})$ is highlighted in bold.}
\label{tab:sim_bdgraph_auc}
\end{table}

\subsection{Simulation 4}

In our fourth simulation study, we evaluated the behavior of each method under a null graph structure, where no interactions are present among the variables. Data were generated under $n = 500$ and $p = 10$, specifically using a mix of 4 continuous variables, 3 count variables, and 3 categorical distributions. Gaussian nodes were sampled from a standard normal distribution, count nodes from a $\text{Poisson}(\mu = 4)$ distribution, and categorical nodes from a balanced two-level multinomial distribution. To assess the robustness of the models to the presence of noise, we also introduced missing values at random in a subset of the data, considering missing rates $\rho_{m} = 0.1$ and $\rho_{m} = 0.2$.

Table \ref{tab:null_graph_fpr_tn} reports the false positive rate (FPR) and the number of true negatives (TN) for each method. As expected, the true positive rate was zero across all methods, so we focused on the FPR as a measure of spurious edge detection. The proposed BMGM method demonstrated excellent control of false positives, with an FPR of 0 and TN = 45 in both missing data scenarios. Similarly, Huge and BDgraph yielded relatively low false positive rates, though they still identified spurious edges in a small number of replicates. In contrast, MGM exhibited a higher FPR, with an average of 0.058 when $\rho_{m} = 0.1$ and 0.047 when $\rho_{m} = 0.2$, corresponding to an average of approximately 2–3 falsely detected edges.

These results confirm that our proposed BMGM model does not suffer from elevated false positive rates under a true null graph, even when data are missing at random. This is an important property in practice, as real datasets often contain many weak or no associations, and a model prone to detecting noise as signal can mislead inference.

\begin{table}[ht]
\centering
\caption{Simulation 4: Null graph scenario — False positive rate (FPR) and true negatives (TN). Mean (SD) across $K=20$ replicates for $p = 10$, $n = 500$.}
\begin{tabular}{@{}ccllll@{}}
\toprule
$\rho_m$ & Metric & BMGM & BDgraph & huge & MGM \\ \midrule
\multirow{2}{*}{0.1} 
& FPR & \textbf{0 (0)} & 0.007 (0.013) & 0.002 (0.007) & 0.058 (0.051) \\
& TN  & \textbf{45 (0)} & 44.700 (0.571) & 44.900 (0.308) & 42.400 (2.280) \\ \midrule
\multirow{2}{*}{0.2} 
& FPR & \textbf{0 (0)} & 0.008 (0.011) & 0.001 (0.005) & 0.047 (0.044) \\
& TN  & \textbf{45 (0)} & 44.650 (0.489) & 44.950 (0.224) & 42.900 (1.997) \\
\bottomrule
\end{tabular}
\label{tab:null_graph_fpr_tn}
\end{table}

\subsection{Sensitivity Analysis}

In this section, we analyze the sensitivity of the proposed model to the choice of the hyperparameters $v_0$, $v_1$, and $\pi$ of the prior \eqref{eq:prior_spikeslab1} on the precision matrix $\Omega$, which determine the prior probability of edge inclusion in the graph. We assess the effects of different hyperparameter settings using simulated data from the scenario presented in Section \ref{sec:random_graph}, considering the full data case with $p = 12$ and $n = 200$.

\begin{figure}
    \centering
\includegraphics[width=0.4\textwidth]{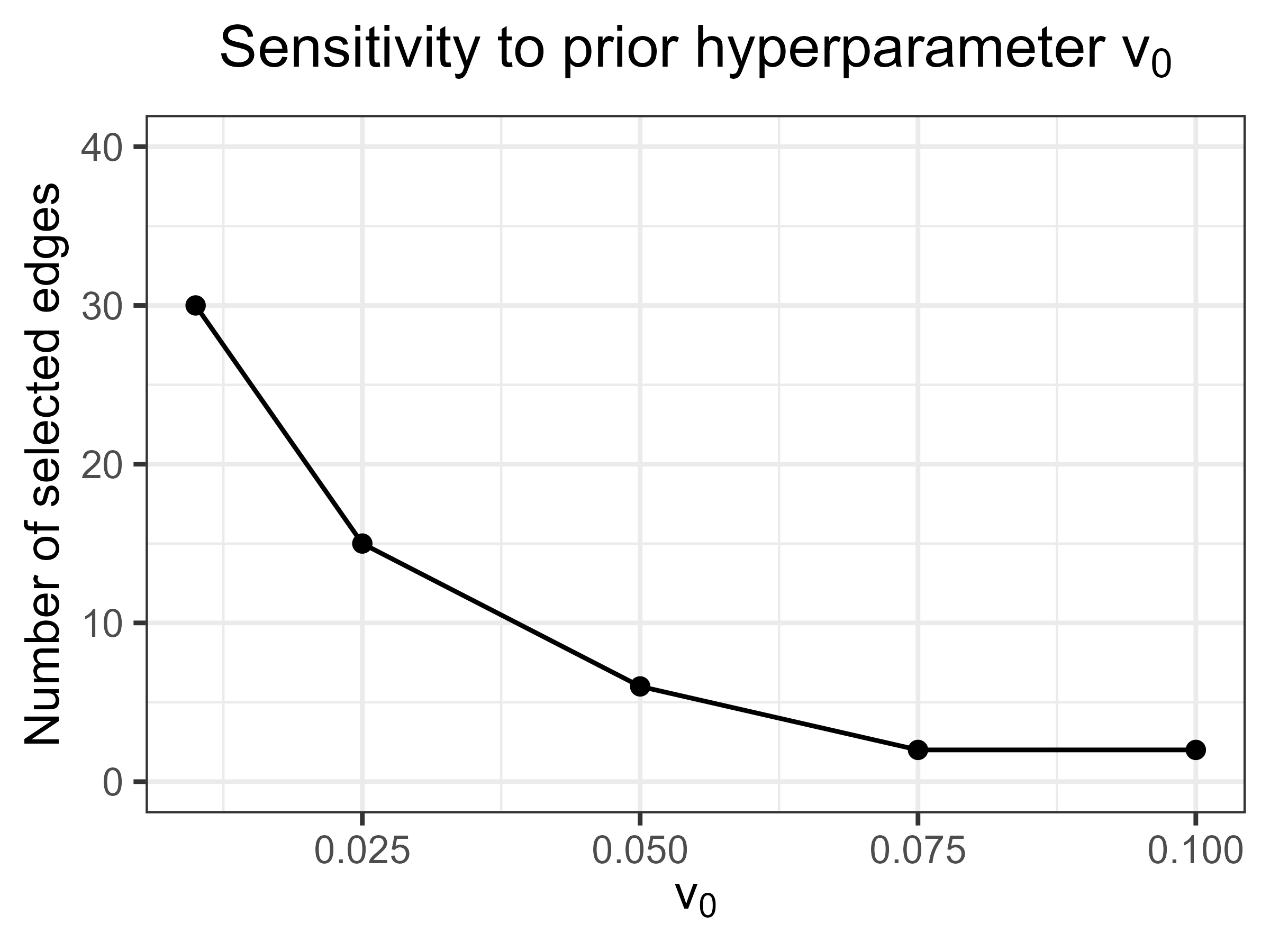}
  \includegraphics[width=0.4\textwidth]{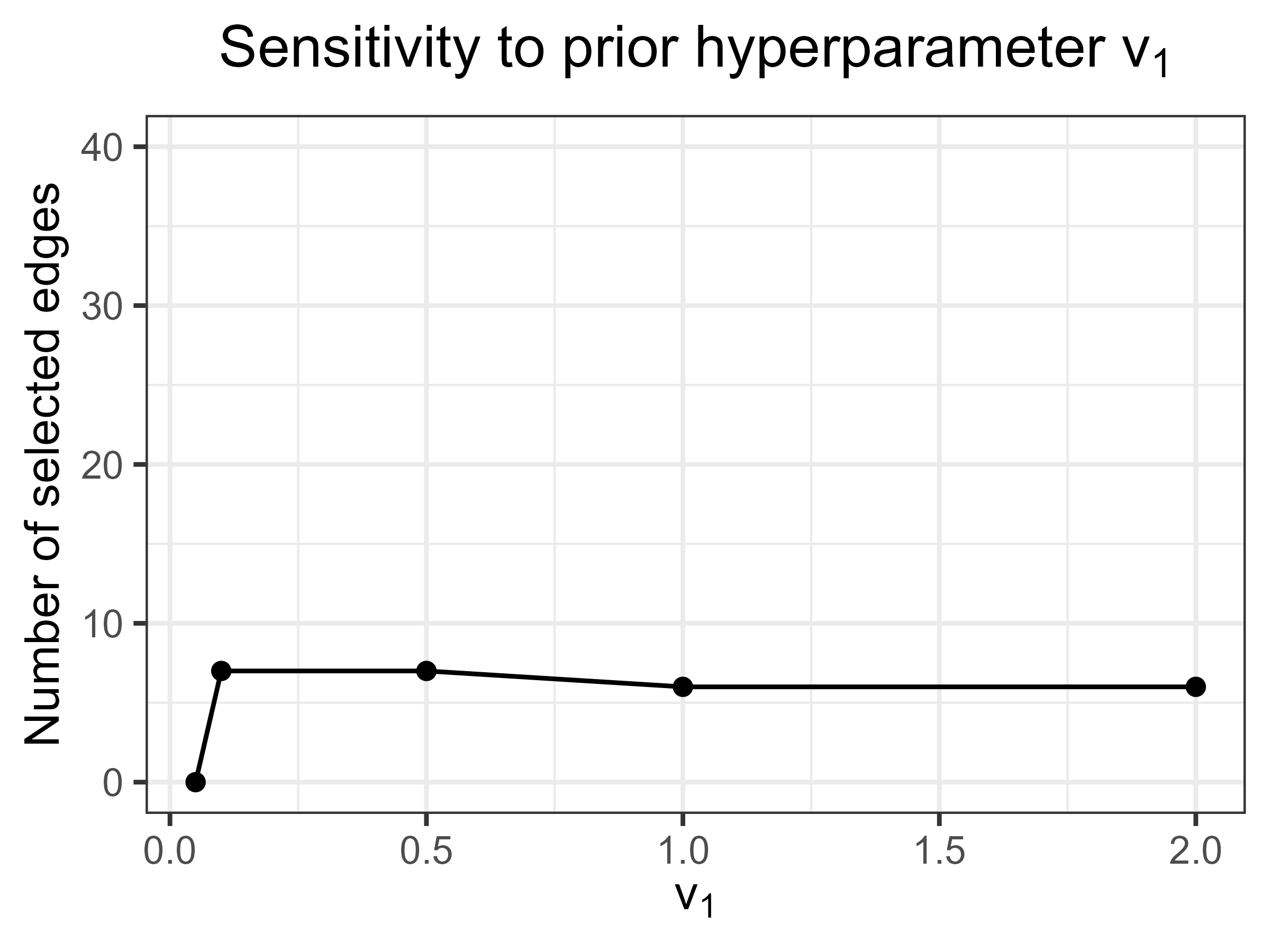}
   \caption{Left: Effect of varying $v_0$ on the number of selected edges, with $v_1=1$ and $\pi = \frac{2}{p-1}$. Right: Effect of varying $v_1$ on the number of selected edges, with $v_0=0.05$ and $\pi = \frac{2}{p-1}$.}
            \label{fig:sens_v0v1}
                   \end{figure}

\begin{figure}
\centering
        \includegraphics[width=0.4\textwidth]{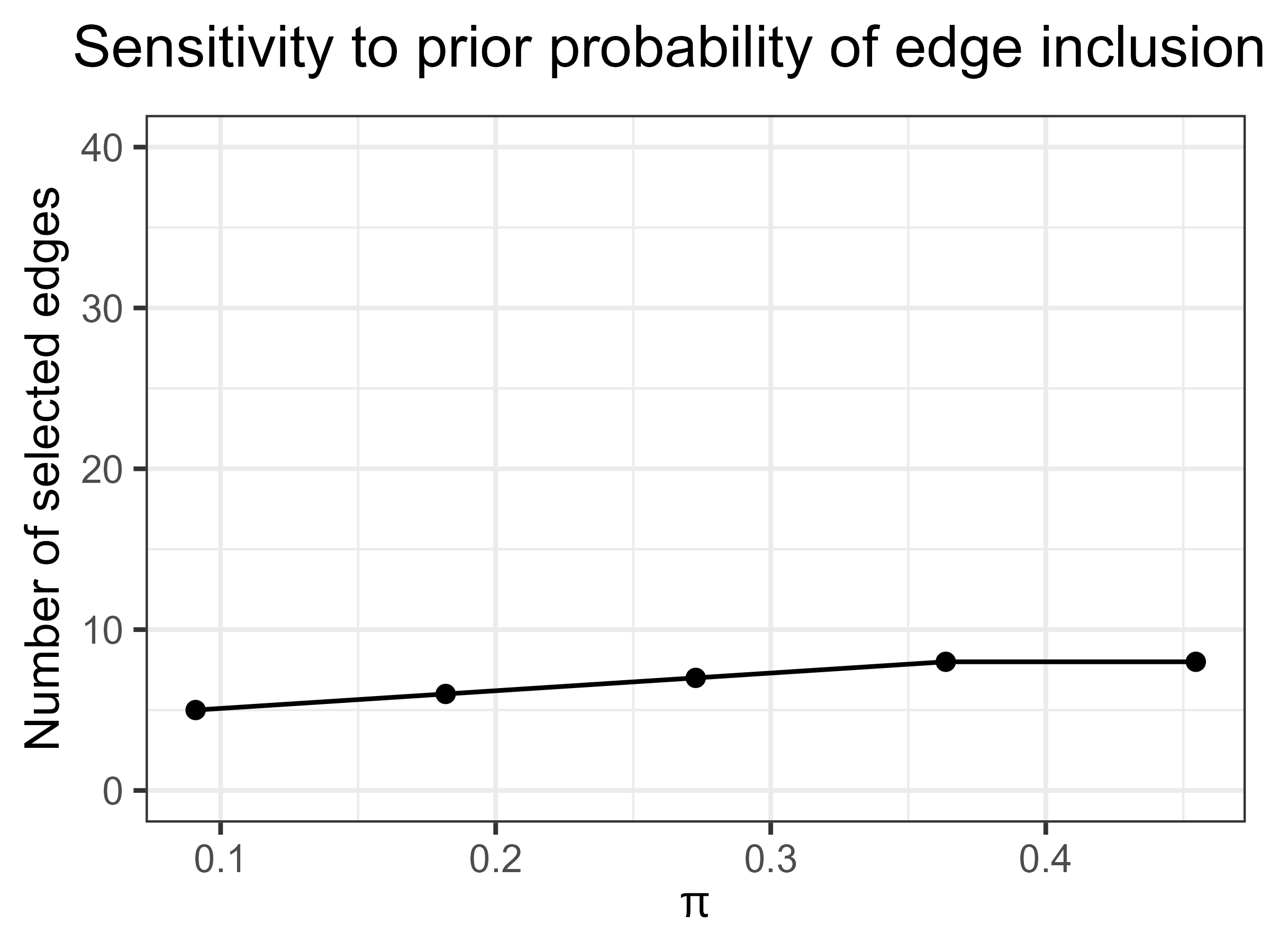}
     \includegraphics[width=0.4\textwidth]{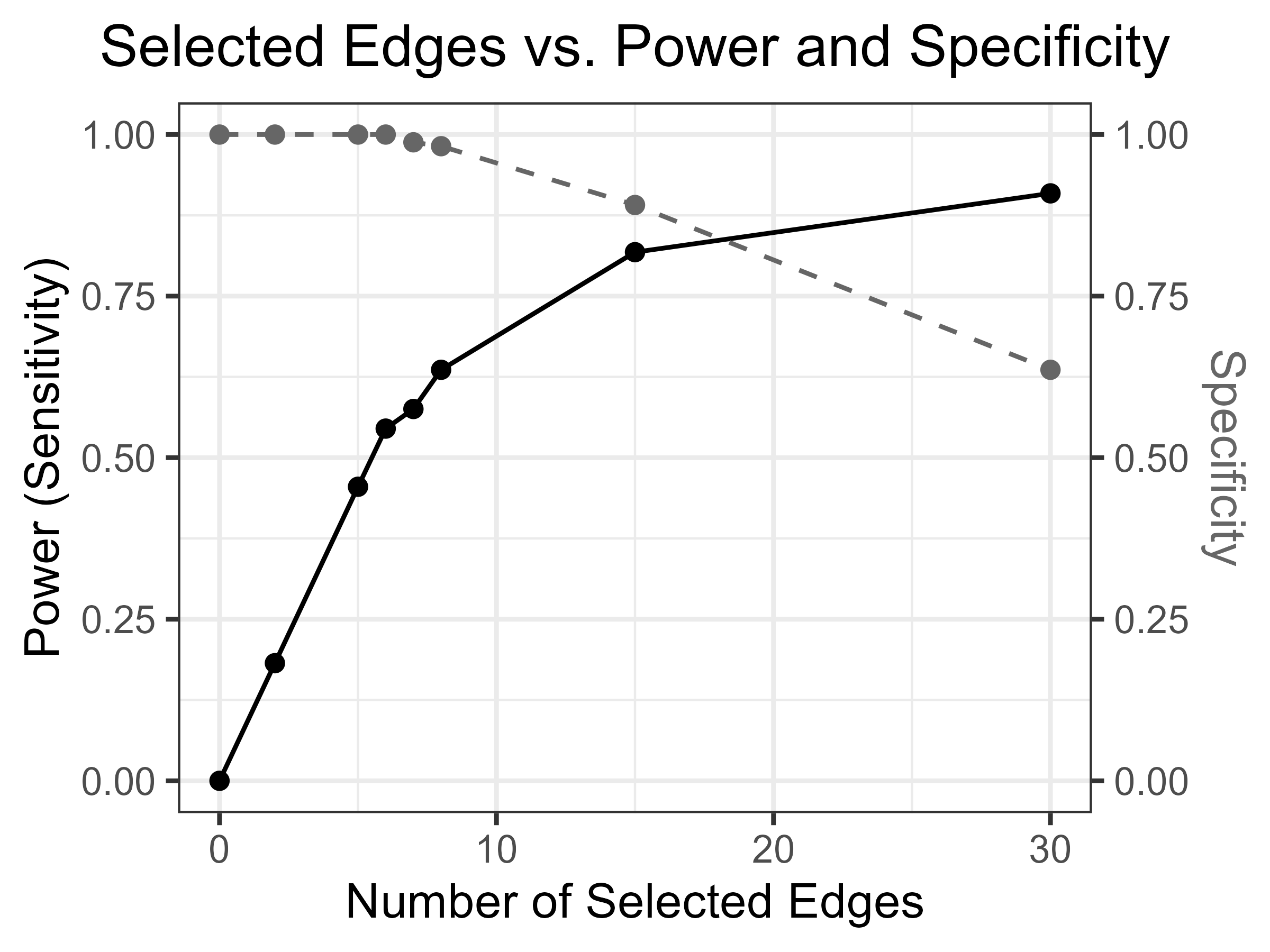}
             \caption{Left: Effect of varying $\pi$ on the number of selected edges, with $v_0=0.05$ and $v_1=1$. Right: Sensitivity (solid line) and specificity (dashed line) of the model across the number of selected edges, aggregated over all prior settings.}
        \label{fig:sens_num_edges}
\end{figure}

Figure \ref{fig:sens_v0v1} (left plot) illustrates the effect of $v_0$ on the edge inclusion. In this case, $v_1$ is fixed at $1$ and $\pi$ is set to $\frac{2}{p-1}$. It is evident that when $v_0$ is set to a very small value, there is a notable increase in the number of selected edges. This occurs because, in this case, the prior for $\beta_{ij}$ tends to be similar to a point-mass mixture that selects any $\beta_{ij} \neq 0$ as an edge. Conversely, Figure \ref{fig:sens_v0v1} (right plot) depicts the effect of $v_1$. In this case, $v_0$ and $\pi$ are set as $0.05$ and $\frac{2}{p-1}$, respectively. We observe that when both $v_1$ and $v_0$ are set to 0.05 the mixture collapses, resulting in the model detecting no edges. However, across the other values considered, $v_1$ does not appear to significantly influence the number of edges identified in the graph, suggesting a degree of flexibility in the choice of $v_1$. Finally, Figure \ref{fig:sens_num_edges} (left plot) indicates that results appear to be relatively insensitive to the value of $\pi$ within the range examined when $v_0 = 0.05$ and $v_1 = 1$.  These  findings are consistent with those of \citet{wang2015scaling}, who suggests that setting $v_0 \geq 0.01$, $v_1 \leq 10$, and $\pi = \frac{2}{p-1}$ typically yields satisfactory performance.  Additionally, Figure \ref{fig:sens_num_edges} (right plot) illustrates the trade-off between sensitivity and specificity as the number of selected edges increases across all prior settings.

 %   \caption{Prior sensitivity analysis: Effects of the hyperparameters $v_0$, $v_1$, and $\pi$ on the edge inclusion in the graph.}
   % \label{fig:two_figures}
%\end{figure}

\section{Application to Eating Disorder Symptoms}
\label{sec:application}
We illustrate the application of the proposed model to real data collected from adolescents diagnosed with an eating disorder (ED). This study was determined to be exempt by the UTSW Institutional Review Board. 

\subsection{Participants and Data}
Data was obtained from 313 adolescents, 237 diagnosed with Anorexia Nervosa, and 40 with Bulimia Nervosa. All participants were adolescents aged 12-18 years old (M = 14.57, SD = 1.56) who were admitted for inpatient or partial hospitalization treatment at a large pediatric eating disorder center from 2018 to 2023. %The majority of the sample were adolescent girls (N = 345), identified as white (48.6\%), and were diagnosed with anorexia nervosa (66.4\%).% 
Adolescents completed assessments at intake and again at discharge. % from the ED treatment using REDCap, a secure web application for building and managing online surveys and databases. As data was originally collected for treatment purposes and de-identified for this research, the UT Southwestern Medical Center IRB determined that the study was exempt and did not require consent of participants. 
Demographic information (age, sex, diagnosis) was collected at intake from the parent or guardian or the medical chart. 

ED symptoms were assessed using the Eating Disorder Examination Questionnaire (EDE-Q) \citep{fairburn1994}. This questionnaire includes 22 items with responses that are restricted to a 7-point Likert-type scale (0-6) and six free-response items that allow writing in a numerical response to queries about the number of times or days that specific disordered eating behaviors occurred in the last month. Conventional scoring of the EDE-Q yields subscale scores for eating restraint (EDE-Q:R; average of 5 items), eating concern (EDE-Q:E; 5 items), shape concern (EDE-Q:S; 8 items), and weight concern (EDE-Q:W; 5 items) and a total score that is the average of the 4 subscale scores. None of the subscale scores include the information obtained from the six free-response items, which are clinically relevant as they provide a measure of actual disordered eating behavior occurring in the individual. For example, a patient’s worries about loss-of-control eating is part of the EDE-Q:E subscale but actual episodes in the last month are only assessed in the free-response item. Variables are described more in detail in Table \ref{tab:descriptive}.

\begin{table}
\centering
\begin{tabular}{lcc}
\toprule
 & Intake ($n = 301$)& Discharge ($n = 286$) \\  \midrule
   \textbf{Categorical} & \multicolumn{2}{c}{Proportions} \\
\cline{2-3} 
  Sex & F = 0.89, M = 0.11 & F = 0.88, M = 0.12 \\ 
  Diagnosis & AN = 0.88, B = 0.12 & AN = 0.87, B = 0.13 \\  \midrule
  \textbf{Zero-Inflated counts} & \multicolumn{2}{c}{\% Zero - Mean (SD) (Non-Zero)} \\
\cline{2-3} 
  EDE-Q:13 & 64.5\% - 7.82 (8.30) & 80.4\% - 12.46 (12.17) \\ 
  EDE-Q:14 & 62.3\% - 7.82 (7.97) & 78.0\% - 12.73 (15.17) \\ 
  EDE-Q:15 & 69.8\% - 7.36 (7.38) & 84.2\% - 4.33 (4.89) \\ 
  EDE-Q:16 & 72.3\% - 11.33 (10.83) & 83.9\% - 4.43 (5.13) \\ 
  EDE-Q:17 & 90.4\% - 6.69 (7.88) & 95.5\% - 3.77 (4.32) \\ 
  EDE-Q:18 & 41.3\% - 2.98 (1.92) & 69.5\% - 1.77 (1.43) \\ \midrule
\textbf{Continuous} & \multicolumn{2}{c}{Mean (SD)} \\
\cline{2-3} 
Age & 14.57 (1.56) & 14.6 (1.59) \\ 
  EDE-Q:R & 3.40 (2.06) & 1.81 (1.80) \\ 
  EDE-Q:E & 2.70 (1.73) & 1.93 (1.58) \\ 
  EDE-Q:S & 4.06 (1.96) & 3.70 (2.05) \\ 
  EDE-Q:W  & 3.49 (2.01) & 3.11 (2.03) \\ 
  BMI-z & -1.27 (1.75) & -0.39 (0.94) \\ 
\bottomrule
\end{tabular}
\caption{Application to eating disorder symptoms. Participants characteristics and types of variables considered}
\label{tab:descriptive}
\end{table}

Below we report results obtained by analysing the data at intake and discharge, separately. From the 313 total participants, at intake we considered 301 since 12 subjects had atypical values or no responses in all the questions of the EDE-Q. Similarly, at discharge we considered 286 as 26 did not have any response in the questionnaires, and 1 had atypical values. Both at intake and discharge, we had 16 kids with one or two variables missing. These were included in the analyses, since our methodology is able to handle missing data. 

\subsection{Parameter settings and convergence diagnostics}
Results we report here were obtained under non-informative priors, as described in Section  \ref{sec:simsettings}, and by running the MCMC for 10,000 iterations, with 5,000 burn-in. To monitor convergence, we calculated the multivariate potential scale reduction factor (MPSRF - $\hat{R}$) of \cite{brooks1998general} using 3 chains. We also monitored the trace plots of the edge-potential parameters. At intake, the estimated $\hat{R}$ was 1.05, below the suggested threshold of 1.2 \citep{brooks1998general}. Also, the univariate potentials scale reduction factors calculated for all elements were below 1.08. This highlights the strong convergence in this scenario. %Figure \ref{fig:tp_intake} depicts the trace plot of the edge-potential parameters in the intake case.
Similarly, at discharge, the MPSRF - $\hat{R}$ of the edge-potential parameters was equal to 1.06, and the maximum of the univariate PSRF was 1.04, suggesting strong convergence across the chains. 
%Trace plots of the posterior edge-potentials can be found in Figure \ref{fig:tp_discharge}.

\subsection{Results}
\begin{figure}
    \centering
        \includegraphics[width=0.9\linewidth]{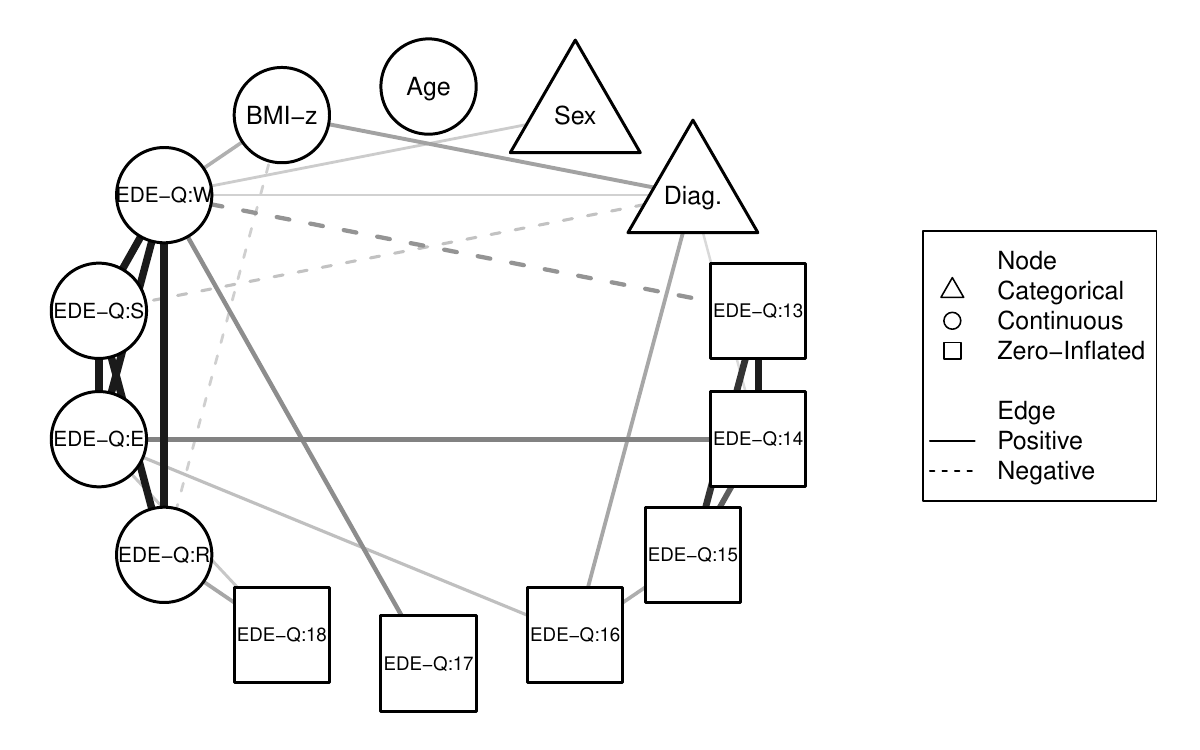}\\
           \includegraphics[width=0.9\linewidth]{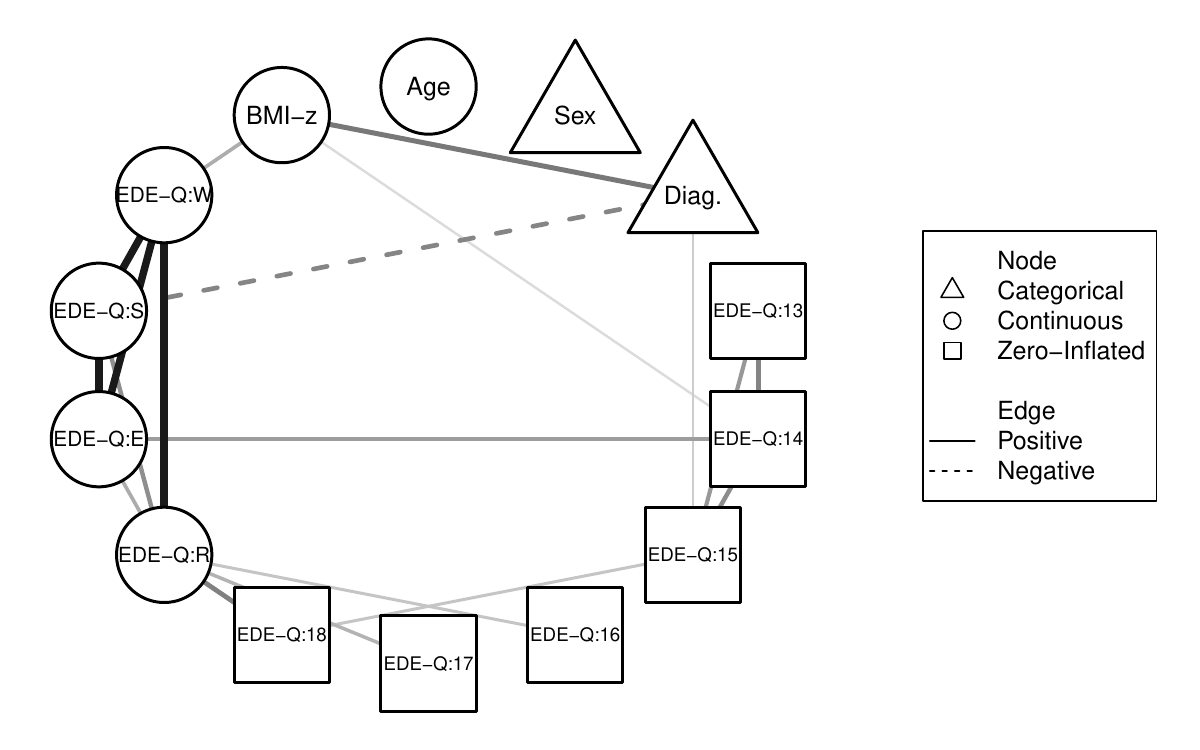}
        % No caption or label for the legend
    \caption{Application to eating disorder symptoms: Inferred graph at intake (top) and at discharge (bottom).  Nodes are represented by geometric figures; Triangles correspond to categorical variables, circles represent continuous variables and zero-inflated counts variables are symbolized by squares. Positive interactions are represented by continuous lines, and negative interactions by dashed lines. The width of the edge between two nodes is proportional to the magnitude of the corresponding estimated element of the precision matrix.}
    \label{fig:combined}
\end{figure}

Figure \ref{fig:combined} shows the graph inferred at intake (top) and at discharge (bottom). At intake, we observe a positive interaction between diagnosis and BMI-Z score, which is expected as the diagnostic criteria for anorexia nervosa in adolescents require weight loss or a failure to gain weight appropriately, while the diagnostic criteria for bulimia nervosa do not require weight changes. Similarly, there is a negative interaction between EDE-Q:S and diagnosis, suggesting that adolescents diagnosed with bulimia nervosa report fewer concerns about their shape than those diagnosed with anorexia nervosa. Another negative interaction observed is between EDE-Q:W and EDE-Q:13, suggesting that higher concerns about weight are associated with engaging in fewer episodes of over-eating. The three free-response queries related to binge-eating behaviors (EDE-Q:13 (eating a large amount), EDE-Q:14 (loss-of control eating), EDE-Q:15 (days of overeating and loss of control)) are closely associated with each other. In general, these symptom relationships are expected and consistent with the diagnoses.

At discharge, we note that the structure of the interactions remains similar. However, certain interactions were not observed or disappeared compared to the graph at intake. For instance, for the EDE-Q:W (weight concern), both its negative association with EDE-Q:13 (eating a large amount) and the positive association with Sex and the diagnosis are no longer present at discharge.  This suggests that the treatment process may have reduced these relationships, a result expected as the treatment process reduces the cognitive distortions related to weight (EDE-Q:W), diminishes excessive eating (EDE-Q:13), and attenuates differences in symptoms across diagnosis and sex. 

Several other estimated dependencies are of clinical interest. At intake, exercise (EDE-Q:18) is associated only with the restraint subscale (EDE-Q:R) and the eating concern subscale (EDE-Q:E), while vomiting behaviors (EDE-Q:16) is associated with eating concerns (EDE-Q:E), and laxative use (EDE-Q:17) is associated with weight concerns (EDE-Q:W). In contrast, at discharge, exercise (EDE-Q:18) has become associated with restraint concerns (EDE-Q:R) as well as the number of days of overeating/loss of control eating (EDE-Q:15). In addition, both vomiting and laxative use are now related to the restraint subscale. These changes imply that, at discharge, vomiting (EDE-Q:16) became conditionally independent of the eating concerns given the value of the restraint subscale (EDE-Q:R). Similarly, at discharge laxative use (EDE-Q:17) became conditionally independent of weight concerns (EDE-Q:W), given the restraint scale (EDE-Q:R).
Finally, it is noteworthy that age appears independent of all other variables considered, both at intake and discharge, but at discharge, both age and sex are independent from all the other variables. This suggests that symptoms among adolescents affected by ED do not vary with age or sex after treatment, with a limitation being that the sample is dominated by female adolescents.

These changes in relationships between the more cognitive and behavioral measures of disordered eating may reflect the ability of adolescent patients to engage in behaviors during intensive treatment, shifts in weight that alter behaviors, or changes in cognitive features of eating disorders in response to treatment. A better understanding of how cognitions and weight support behaviors at different stages of eating disorder treatment in adolescents may be helpful in reducing relapse and persistence. Relapse and persistence of eating disorders into adulthood is common, with a recent meta-analysis suggesting $40\%$ persist and $25\%$ relapse after achieving recovery \citep{filipponi2022}.

To address possible concerns that differences between the intake and discharge graphs could be driven by changes in the sample composition, we repeated the analysis using the subset of 283 individuals with complete data at both time points. The resulting graphs were nearly identical to those shown in Figure \ref{fig:combined}, suggesting that the observed differences are not explained by differences in the individuals included in each cohort.

\section{Conclusion}
\label{sec:conclusion}

In our paper, we introduce a Bayesian pairwise mixed graphical model that effectively models interactions between different types of variables, such as continuous, discrete, categorical, and zero-inflated count data. Our approach uses a node-conditional estimation framework and integrates the edge-potential function proposed by \citet{roy2020nonparametric} for count data, allowing estimation of both positive and negative interactions. This flexible formulation enables our model to capture the distribution of individual nodes of different types. It also facilitates the imputation of missing data, by considering potential interactions with other variables. 

We have assessed performances on two simulated scenarios with distinct dependence structures, that also include cases with data missing at random, and compared results with existing methods. Results have indicated that the proposed model is robust to missing data and effectively captures both positive and negative interactions between mixed types of data. Furthermore, performances of our model have matched or in some cases outperformed existing methods, especially in scenarios with missing data or when including zero-inflated count data. Finally, we have illustrated the ability of our model to analyze real data from adolescents diagnosed with an eating disorder, where nearly half of the variables are zero-inflated and with some variables containing missing data. Model fitting has resulted in estimated sparse graphs that have shown some expected interactions, along with others that unveil intrinsic relationships. For instance, at discharge, previous connections between weight concerns (EDE-Q:W) with EDE-Q:17 (laxative use), EDE-Q:13 (eating a large amount), and the BMI-z score disappeared, implying that the treatment may have contributed to reducing cognitive distortions related to weight and unhealthy eating behaviors, as well as supporting weight normalization.
 
In summary, our model offers a flexible framework for modeling dependencies among various types of data, including in the presence of missing values and zero-inflated counts. In the real data application presented in Section \ref{sec:application}, our implementation completes 100 MCMC iterations in approximately 40 seconds. For comparison, BDgraph completes 1,000 iterations in just over one second, while Huge and MGM also converge substantially faster. Despite the added computational cost, our model remains particularly useful in settings where standard methods cannot accommodate structural zeros or missingness.  Furthermore, the MCMC algorithm we have implemented, based on conditional likelihoods and approximations of the normalizing constants, offers a substantial improvement in efficiency, achieving an 83\% reduction in computation time per iteration compared to alternative approaches that rely on the Exchange Algorithm of \cite{murray2012mcmc}, which requires auxiliary variable sampling at every step (see the Appendix). This improvement makes the method considerably more practical for real-world applications. 

\section*{Software and Data Availability}
Code reproducing the results of this paper is available on GitHub at \url{https://github.com/mauroflorez/BMGM} (to be made public upon acceptance of the paper). 
Data used in the application can be obtained from Dr. Carrie McAdams, upon reasonable request.

\section*{Acknowledgments}
We thank the Editor, Associate Editor, and anonymous referees for their thoughtful and constructive feedback, which greatly improved the quality of this manuscript. Data collection was supported by grant NIMH R21 MH131865 (McAdams), CTSA UL1TR003163 from NIH (NCATS) and Children’s Health$^{\text{SM}}$. The content is solely the responsibility of the authors and does not represent official views of NIH or Children’s Health$^{\text{SM}}$. The second author was partially supported by the Italian Ministry of University and Research (MUR), Department of Excellence project 2023-2027 ReDS 'Rethinking Data Science' - Department of Statistics, Computer Science, Applications - University of Florence and the MUR-PRIN grant 2022 SMNNKY, CUP B53D23009470006 European Union - NextGenerationEU, Mission 4 Component 2.

\bibliographystyle{apalike}
\bibliography{sample.bib}       % Bibliography file (usually '*.bib')

\section*{Appendix}

\subsection*{Proof of Proposition 2.1} 
Following \citet{chen2015selection}, we say that two conditional densities are compatible if there exists a function $g$ that is capable of generating both conditional densities. When $g$ is a density, the conditional densities are said to be strongly compatible.

Suppose that for each variable $x_s$, the conditional density is given as in \eqref{eq:conditional}. If $\beta_{ts} = \beta_{st}$, then the conditional densities are compatible. Additionally, any function $g$ that is capable of generating the conditional densities defined in \eqref{eq:conditional} can be expressed as:
\begin{equation}
g(x) \propto \exp\left\{\sum_{s=1}^p f_s(x_s) + \frac{1}{2}\sum_{s=1}^p\sum_{t\neq s} \beta_{ts}F(x_t)F(x_s))\right\}.
\label{eq:g_def}
\end{equation}

\noindent We present a proof similar to the one given in \citet{chen2015selection} based on the work of \citet{besag1974spatial}. First, we can see that $g$ in \eqref{eq:g_def} can also be expressed as 
\begin{equation*}
    \exp\{f_s(x_s) + \frac{1}{2}\sum_{t: t\neq s} (\beta_{st} + \beta_{ts})F(x_t)F(x_s))\}\exp\{\sum_{t \neq s} f_t(x_t) + \frac{1}{2}\sum_{t: t\neq t}\sum_{j: j\neq s, j \neq t}\beta_{tj}F(x_t)F(x_j))\}.
\end{equation*}

\noindent The integrability of $p(x_s | \bm x_{-s})$ guarantees the integrability of $g$ with respect to $x_s$. Therefore, the conditional densities are compatible if $\beta_{ts} = \beta_{st}$. 

Now we have to prove that any function $h$ that is capable of generating the conditional densities is in the form of $g$ as in 
\eqref{eq:g_def}. Let's denote $x^* = F(x)$ and define $P(x^*) = \log\left(\frac{h(x^*)}{h(0)}\right)$. Under this definition, $P(0) = \log\left(\frac{h(0)}{h(0)}\right) = 0$. As pointed out by \citet{besag1974spatial}, there exists an expansion of $P(x^*)$ of the form
\begin{eqnarray}
    P(x^*) = \sum_{s = 1}^p x_s^* G(x_s^*) + \sum_{t\neq s} \frac{1}{2}G_{ts}(x_s^*, x_t^*) x_t^* x_s^*  +
    \sum_{t \neq s, t \neq j, j \neq s} \frac{1}{6}G_{tsj}(x_t^*, x_s^*, x_j^*) x_t^* x_s^* x_j^* + \cdots
\label{eq:expansion}
\end{eqnarray}

\noindent Since we assume $h$ is capable of generating the conditional density $p(x_s | \bm x_{-s})$, we have that
\begin{equation*}
    P(x^*) - P({x_s^0}^*) = \log \left\{ \frac{h(x^*)/\int h(x^*) d_{x_s^*}}{h({x_s^0}^*)/ \int h(x^*) d_{x_s^*}} \right\}   = \log \left \{ \frac{p(x_s | \bm x_{-s})}{p(0 | \bm x_{-s})}\right \},  
\end{equation*}

\noindent where ${x_s^0}^* = (F(x_1),\ldots,F(x_{s-1}), 0, F(x_{s+1}), \ldots, F(x_p))$. It follows that 
\begin{equation*}
    \log \left \{ \frac{p(x_s | \bm x_{-s})}{p(0 | \bm x_{-s})}\right \} = x_s^* ( G(x_s^*) + \sum_{t\neq s} \frac{G_{ts}(x_s^*, x_t^*)}{2} x_t^* + \sum_{t \neq s, t \neq j, j \neq s} \frac{G_{tsj}(x_t^*, x_s^*, x_j^*)}{6} x_t^* x_j^* + \cdots).
\end{equation*}
Letting $x_t^* = 0$ for $t \neq s$ we have 
\begin{eqnarray*}
    x_s^* G_s(x_s^*) = f_s(x_s) - f_s(0).
\end{eqnarray*}
Similarly, if $x_j^* = 0$ for $j \neq t$, $j \neq s$, for the second-order interaction $G_{ts}$, 
\begin{eqnarray*}
    x_s^* G_s(x_s^*) + x_s^* x_t^* G_{ts} (x_t^*, x_s^*) = f_s(x_s) + \beta_{st} F(x_t) F(x_s).
\end{eqnarray*}
Applying the same argument to $P(x^*) - P({x_t^0}^*)$, we obtain
\begin{eqnarray*}
    x_t^* G_t(x_t^*) + x_s^* x_t^* G_{st} (x_s^*, x_t^*) = f_t(x_t) + \beta_{ts} F(x_t) F(x_s).    
\end{eqnarray*}

\noindent Therefore if $\beta_{st} = \beta_{ts}$, then $G_{st}(x_s,x_t) = G_{ts}(x_t,x_s) = \beta_{st}$. We can see that when setting $x_k = 0$, where $k \neq s, k \neq t, k \neq j$, the third-order interactions in \eqref{eq:expansion} are zero. Similarly, higher-order interactions are also equal to zero. Therefore,

\begin{eqnarray*}
    P(x^*) = \sum_{s = 1}^p f_s(x_s) +  \frac{1}{2}\sum_{s=1}^p \sum_{t\neq s}\beta_{ts}F(x_s) F(x_t)
\end{eqnarray*}

\noindent Furthermore, $P(x^*) = \log\{h(x^*))/h(0)\}$, so $h$ is of the form

\begin{eqnarray*}
    h(x^*) = \exp(P(x^*)) = \exp\left\{\sum_{s = 1}^p f_s(x_s) +  \frac{1}{2}\sum_{s=1}^p \sum_{t\neq s}\beta_{ts}F(x_s)F(x_t)\right\}.
\end{eqnarray*}

\noindent Then, the conditional densities are strongly compatible. This means we can arrive at the joint distribution via the conditional densities and that we can infer the parameters of the joint distribution via the conditional densities.

\subsection*{Proof of Proposition 2.2} 

We say that two variables $x_s$ and $x_t$ are conditionally independent if $p(x_s, x_t | x_{-(s,t)}) = p(x_s|x_{-(s,t)})p(x_t|x_{-(s,t)})$, where $x_{-(s,t)}$ are all the random variables except $x_s$ and $x_t$. Following the proof given by \citet{roy2020nonparametric}, we have that the conditional probability between $x_s$ and $x_t$ is given by:
\begin{equation*}
        p(x_t, x_s | x_{-(t,s)}) = \frac{\exp \left(\sum_{h \in \{s,t\}} f(x_h) + \sum_{g \neq h} \beta_{gh}F(x_g)F(x_h) \right)}{\int_S \int_T \exp \left(\sum_{h \in \{s,t\}} f(x_h) + \sum_{g \neq h} \beta_{gh}F(x_g)F(x_h) \right)dT dS}
\end{equation*}

\noindent with $S$ and $T$ the domain space of $x_s$ and $x_t$, respectively. Since $\beta_{ts} = \beta_{st} = 0$, we can break the exponential terms in the previous equation such that it is equivalent to
\begin{equation*}
    \frac{\exp \left(f(x_s) + \sum_{g \neq s} \beta_{gs}F(x_g)F(x_s) \right) \exp\left(f(x_t) + \sum_{g \neq t} \beta_{gt}F(x_g)F(x_t) \right)}{\int_S \exp \left(f(x_s) + \sum_{g \neq s} \beta_{gs}F(x_g)F(x_s) \right) dS \int_T \exp \left(f(x_t) + \sum_{g \neq t} \beta_{gt}F(x_g)F(x_t) \right)dT}
\end{equation*}

\noindent This corresponds to $p(x_s|x_{-(s,t)})p(x_t|x_{-(s,t)})$. Therefore, $x_s$ and $x_t$ are conditionally independent.

\section*{Posterior Inference Using the Exchange Algorithm}

%\subsection{Posterior Sampling}
\label{posterior_sam_exchange_algorithm}
For posterior inference, we also implemented an alternative MCMC algorithm, where we generated proposals via a Gibbs sampler and evaluated them using the Exchange Algorithm proposed by \citet{murray2012mcmc}.  This MCMC algorithm avoids the use of numerical approximations. However, it is computationally more expensive as it requires sampling of auxiliary data.  

The Exchange Algorithm is one extension of the Metropolis-Hastings algorithm to sample from doubly-intractable likelihoods. Specifically, we note that the conditional likelihood at the \textit{s-th} node (\ref{eq:conditional}) can be expressed as 
\begin{equation}
    p(x_s | \bm x_{-s}) = \frac{q(x_s|\bm x_{-s}, \theta_s, {\bm \beta_s})}{A_q(\theta_s, {\bm \beta_s})},
\label{eq:intractable}
\end{equation}
where $q$ represents the unnormalized conditional likelihood and $A_q$ is the normalizing constant that depends on the parameters $\theta_s$ and $\bm \beta_s = \{\beta_{is}\}_{i \neq s}$. We next describe the updates of the individual parameters.

\subsubsection{Node-potential parameters}
We first update $\Theta = (\theta_1, \theta_2, ..., \theta_p)$, the parameters of the node-potential function, based on the following algorithm:
\begin{enumerate}
    \item[i.] For every node $s$, generate a proposal $\theta_s^*$ from a Gaussian distribution centered at the current value $\theta_s$.
    \item[ii.] Sample auxiliary data $x_s^*$ from the likelihood under the proposal $\theta_s^*$. We use an Adaptive Rejection Metropolis Sampler for this step \citep{armspp}.
    
    \item[iii.] Evaluate and accept the new proposal based on the Exchange Algorithm.
    \begin{eqnarray*}
        \alpha(\theta_s, \theta_s^*) &=& \min \left\{ 1, \frac{\prod_{i = 1}^{n} \frac{q(x_{is} | x_{i-s}, \theta_s^*,  \beta_s)}{A_q(\theta_s^*,  \beta_s)}p(\theta_{s}^*) \prod_{i = 1}^{n} \frac{q(x_{is}^* | x_{i-s}, {\theta_s},  \beta_s)}{A_q(\theta_s, \beta_s)}}
        {\prod_{i = 1}^{n} \frac{q(x_{is} | x_{i-s}, \theta_{s},  \beta_s)}{A_q(\theta_{s},  \beta_s)}p(\theta_{s}) \prod_{i = 1}^{n} \frac{q(x_{is}^* | x_{i-s}, \theta_s^*,  \beta_s)}{A_q(\theta_s^*,  \beta_s)}}  \right\}\\
        &=& \min \left\{ 1, \frac{\prod_{i = 1}^{n} q(x_{is} | x_{i-s}, \theta_s^*,  \beta_s) q(x_{is}^* | x_{i-s}, {\theta_s},  \beta_s)p(\theta_{s}^*)}{\prod_{i = 1}^{n} q(x_{is} | x_{i-s}, {\theta_s},  \beta_s)q(x_{is}^* | x_{i-s}, \theta_s^*,  \beta_s)p(\theta_{s})} \right\}
    \end{eqnarray*}
    
where $x_{i-s}$ is the \textit{i-th} observation of all except the \textit{s-th} node.
\end{enumerate}

\subsubsection{Edge-potential parameters}
To update the edge-specific weight parameters, we use the Gibbs sampler proposed by  \citet{wang2015scaling}, which updates each column of the precision matrix $\Omega$ successively. Since the elements $\Omega_{st} = \Omega_{ts} = \beta_{st}$, and the diagonal entries do not change over iterations, updating the \textit{l-th} column of $\Omega$, ${\Omega_l}$, corresponds to update the edge-weight parameters $\bm \beta_l = \{\beta_{il}\}_{i \neq l}$. We let $\Omega_{-l-l}$ denote the submatrix of $\Omega$ removing the \textit{l-th} row and column, and consider $G$ the gram matrix of $F(X)$, $G = (F(x) - \bar{F}(x))^T(F(x) - \bar{F}(x))$. Thus, we update each column $\Omega_l$, $1 \leq l \leq p$, of $\Omega$ successively as follows:

\begin{itemize}
    \item[i.] Generate an update ${\Omega_l}^*$ for $\Omega_l$. A candidate is generated from a MVN$(-C G_{l,-l}, C)$, where $C = (G_{ll}\Omega_{-l-l}^{-1} + D_l^{-1})^{-1}$, with $D_l$ the prior variance corresponding to $\Omega_{l,-l}$. 
    
    \item[ii.] Sample auxiliary data for the node $l$, ${x_l}^*$, from the conditional likelihood under the proposal for the edge-weights ${\Omega_l}^* = {\bm \beta_l}^* = \{{\beta_{il}}^*\}_{i \neq l}$.

    \item[iii.] Evaluate and accept the candidate based on the Exchange Algorithm.
    \begin{eqnarray*}
        \alpha( \bm \beta_l,  {\bm \beta_l}^*) &=& \min \left\{1, \frac{\prod_{i = 1}^{n} \frac{q(x_{il} | x_{i-l},  \theta_l,  \bm \beta_l^*)}{A_q( \theta_l,  \bm \beta_l^*)}p( \bm \beta_{l}^*) \prod_{i = 1}^{n} \frac{q(x_{il}^* | x_{i-l},  \theta_l,  \bm \beta_l)}{A_q( \theta_l,  \bm \beta_l)}}
        {\prod_{i = 1}^{n} \frac{q(x_{il} | x_{i-l},  \theta_l,  \bm \beta_l)}{A_q( \theta_l,  \bm \beta_l)}p( \bm \beta_l) \prod_{i = 1}^{n} \frac{q(x_{il}^* | x_{i-l},  \theta_l,  \bm \beta_l^*)}{A_q( \theta_l,  \bm \beta_l^*)}}  \right\}\\
       &=& \min \left\{1, \frac{\prod_{i = 1}^{n} q(x_{il} | x_{i-l},  \theta_l,  \bm \beta_l^*) q(x_{il}^* | x_{i-l},  \theta_l,  \bm \beta_l)p(\bm \beta_{l}^*) }
        {\prod_{i = 1}^{n} {q(x_{il} | x_{i-l},  \theta_l,  \bm \beta_l)} q(x_{il}^* | x_{i-l},  \theta_l,  \bm \beta_l^*)p( \bm \beta_l)}  \right\}
    \end{eqnarray*}
    
    \item[iv.] Sample the indicators $z_{ij}$. As pointed out by \citet{wang2015scaling}, priors \eqref{eq:prior_spikeslab1} and \eqref{eq:prior_spikeslab2} imply that the $z_{ij}$'s are distributed as independent Bernoulli distributions, with probability
    \begin{equation}
        Pr(z_{ij} = 1 |  \beta,  x) = \frac{N(\beta_{ij}| 0, v_1^2)\pi}{N(\beta_{ij}| 0, v_1^2)\pi + N(\beta_{ij}| 0, v_0^2)(1-\pi)}.
    \end{equation}
\end{itemize}

\end{document}